\documentclass[twocolumn]{article}
\usepackage[english]{babel}
\usepackage{blindtext}
\usepackage{xspace}
\usepackage{rotating}
\usepackage{subcaption}
\usepackage[hyphens]{url}
\usepackage{amsmath}
\usepackage{cleveref}
\crefformat{section}{\S#2#1#3} 
\crefformat{subsection}{\S#2#1#3}
\crefformat{subsubsection}{\S#2#1#3}  
\usepackage{breqn}

\newcommand{\dc}{data center\xspace}
\newcommand{\dcs}{data centers\xspace}

\newcommand{\eg}{\textit{e}.\textit{g}.}

\newcommand{\thss}{\textsuperscript{th}}
\newcommand{\millis}{$\mathrm{ms}$\xspace}
\newcommand{\micros}{$\mu\mathrm{s}$\xspace}

\date{}
\begin{document}
\setlength{\abovedisplayskip}{0pt}
\setlength{\belowdisplayskip}{0pt}
\title{No Delay: Latency-Driven, Application Performance-Aware, Cluster Scheduling}

 \author{Diana Andreea Popescu \qquad \qquad \qquad \qquad \quad Andrew W. Moore \\\quad diana.popescu@cl.cam.ac.uk \qquad \qquad \qquad andrew.moore@cl.cam.ac.uk \\ \quad \quad University of Cambridge \qquad \qquad\qquad \qquad University of Cambridge}

\maketitle
\begin{abstract}
Given the network latency variability observed in data centers, applications' performance is also determined by their placement within the data centre. We present NoMora, a cluster scheduling architecture whose core is represented by a latency-driven, application performance-aware, cluster scheduling policy. The policy places the tasks of an application taking into account the expected performance based on the measured network latency between pairs of hosts in the data center. Furthermore, if a tenant's application experiences increased network latency, and thus lower application performance, their application may be migrated to a better placement.
Preliminary results show that our policy improves the overall average application performance by up to 13.4\% and by up to 42\% if preemption is enabled, and improves the task placement latency by a factor of $1.79\times$ and the median algorithm runtime by $1.16\times$ compared to a random policy on the Google cluster workload. This demonstrates that application performance
can be improved by exploiting the relationship between network latency and application performance, and the current network conditions in a data center, while preserving the
demands of low-latency cluster scheduling.

\end{abstract}
\section{Introduction}

Cloud computing has revolutionised the way businesses use computing infrastructure. Instead of building their own \dcs, companies rent computing resources from cloud providers (\eg, Amazon AWS, Google Cloud Platform, or Microsoft Azure), and deploy their applications on cloud provider hardware. Network latency variability is still common in multi-tenant \dcs~\cite{wang:2010,barker:2010,bobtail, mogul:2015, diana-tma}, and even small amounts of delay, in the order of tens of microseconds, may lead to significant drops in application performance~\cite{pam2017, diana-report}.
An important factor in achieving predictable application performance is understanding the networking requirements of the application in terms of bandwidth and latency. Once these requirements have been determined, they have to be incorporated into the \dc management stack. This can be done in-network, through scheduling~\cite{Hedera, MicroTE, fastpass} or prioritising~\cite{qjump, pfabric, task, varys} the application's flows, and-or at the end-host, through bandwidth allocation~\cite{bwe}. In these situations, the placement of the application's tasks is assumed to be known before incorporating its network resource demands. If the tasks' placements are not known a priori or if they can be changed, the network resource demands can be incorporated at a higher level in the \dc management stack, namely in the cluster scheduler.
Previous work (\cref{sec:bk-cluster-scheduler}) has looked at providing network bandwidth and tail latency guarantees, and, as a result, the application would meet its performance guarantees. However, none of the cluster schedulers have placed an application's tasks according to their expected performance as predicted by the current network conditions. 
In this work, we change the viewpoint: if the tenant wants a certain performance for their application, what network conditions does the application need in terms of latency? If we know how the application reacts to latency and the current network conditions, then we can place the tenant's application in the \dc ensuring the best performance achievable under the current network conditions. Furthermore, network bandwidth demands could be incorporated in the placement decision, or previous systems (\cref{sec:bk-cluster-scheduler}) can be used to meet them.

The relationship between network latency and application performance can help cloud customers to determine the performance their application can achieve under certain network conditions and can guide cloud operators in selecting the network latency ranges that best suit the needs of their customers. By measuring dynamically the network latency in the \dc and having a model of the application performance dependent upon network latency, the expected application performance under the measured network conditions can be determined. 

In this work, we show how to build functions that predict application performance based upon network latency for selected cloud applications (key-value store and machine learning frameworks) from experimental data. These functions are then abstracted in a way in which they can be understood by a cluster scheduler. We use these functions in a cluster scheduling architecture, NoMora\footnote{\emph{mora} means delay in Latin, so the name refers to applications being scheduled to not have network delay affecting their performance}, that extends the Firmament~\cite{firmament} cluster scheduling framework. The core of NoMora is a cluster scheduling policy that places the tasks of an application (job) taking into account the expected performance based on the measured network latency between pairs of hosts in the data center. Furthermore, if a tenant's application experiences increased network latency, and thus lower application performance, the application may be migrated to a different host. 

A commonality of the cluster workloads released by companies is that they do not include information related to networking demands, \eg, network bandwidth, or how latency-sensitive the application is. 
Therefore, this lack of information represents a challenge when developing new cluster scheduling policies that try to improve application performance while considering networking demands. To overcome this challenge, we augment the traces with application performance predictions dependent upon network latency determined experimentally in Section~\ref{sec:impact-model}. This represents a first step towards a comprehensive cluster workload which includes not only the usual information (CPU, RAM, etc.), but also information related to network resources required, thus offering a full picture of the applications deployed.

In this paper, we make the following contributions:
\begin{itemize}
\item How to build functions that predict application performance determined from experimental data and a way to abstract and map these functions to a cluster scheduling policy.
\item A cluster scheduling architecture NoMora that uses these performance functions and measurements of network conditions, and whose core is represented by
a latency-driven, application performance-aware, cluster scheduling policy.
\item An evaluation of our policy using well-known cluster workloads augmented with our performance predictions and network latency measurements conducted in previous work~\cite{diana-tma}, showing notable improvement in the overall average application performance compared to the baselines. 
\end{itemize}

\section{Motivation}
\label{sec:motivation}

\begin{figure}
 \centering
    \begin{minipage}{0.4\textwidth}
        \centering
        \includegraphics[width=0.98\textwidth]{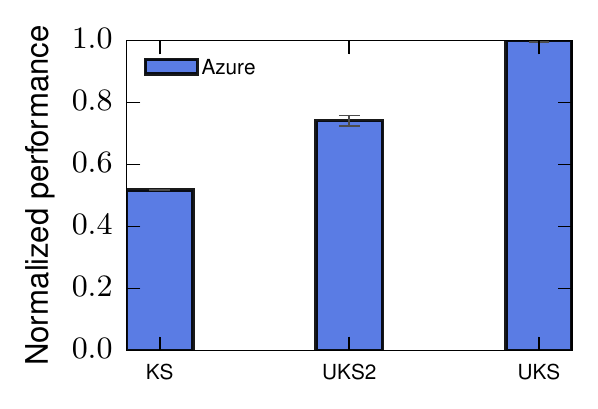}
       \subcaption{Normalised performance}
       \label{memcached-azure-perf}
    \end{minipage}
\centering
   \begin{minipage}{0.35\textwidth}
        \centering
        \includegraphics[width=0.98\textwidth]{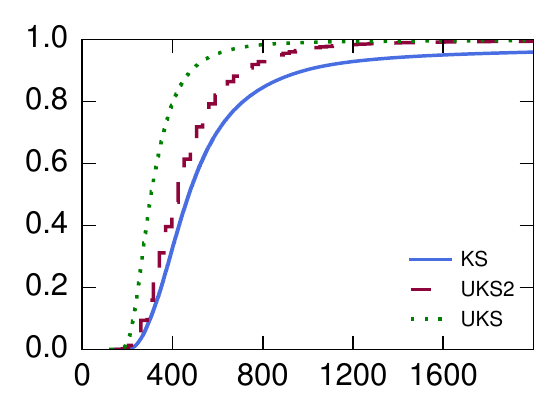}
        \subcaption{Request-response latency (\textbf{\micros})}
        \label{memcached-azure-lat}
    \end{minipage}
 \caption{Running one Memcached server and four clients in two different zones and at different times within the same zone gives different performance.}
\label{fig:memcached-azure}
\end{figure}

We run an experiment with a Memcached server running in one VM, and four clients running on five different VMs in one cloud provider (Microsoft Azure). The clients send requests to the server using the Mutilate~\cite{mutilate} benchmark for a period of time, and we measure the number of requests per second reported by the benchmark. We perform this experiment ten times using the same setup (same specifications for hosts and the same software), but in three different settings. The VMs were deployed in the Korea South zone (KS), then in the UK South zone on a Sunday evening, when there the network is less utilized~\cite{diana-tma}), and again in the UK South zone during the day on Monday after restarting the VMs. Figure~\ref{fig:memcached-azure} shows the performance obtained in the three cases normalised with respect to the maximum obtained performance (in UK South during Sunday evening). The performance obtained differs between the three different scenarios, and even within the same zone, with less than 80\% of the performance achieved in the second scenario, and with only 50\% performance achieved in the first scenario. There is a performance difference depending on the placement of the application and the current network conditions in the \dc. In this situation, knowing how the application reacts to network conditions would guide the placement within the \dc. If the application is already running and the network conditions change, reducing the application's performance, then it can be migrated to a different placement. 

\begin{figure}
\centering
\includegraphics[scale=1]{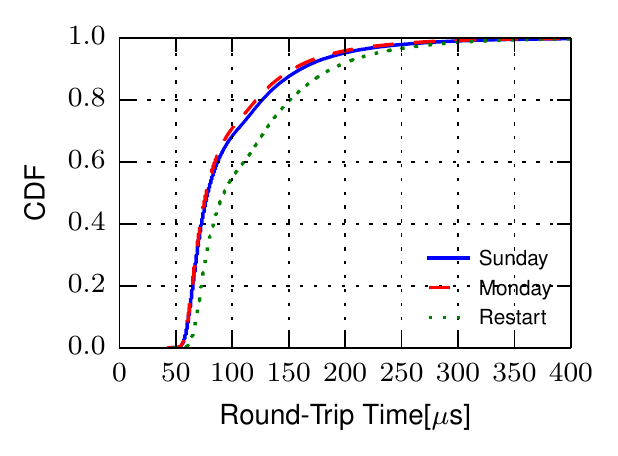}
\caption{RTT between 2 VMs at different times during the day and after restarting the VMs.}
\label{fig:rtt-gcp}
\end{figure}

In a different experiment, we observed that network latency is not constant between VMs, as can be seen from Figure~\ref{fig:rtt-gcp} during an experiment run during the week of 27th of January 2019. In this experiment we measured the Round-Trip Time (RTT) overal several hours (10 million measurements) using UDP probes between two basic type VMs rented from a different cloud provider (GCP). The first two experiments use the same VMs during different periods of time, and as can be seen the results are similar. For the third run, we restarted the VMs, and got different latencies than in the previous two cases.

\section{Predicting application performance}
\label{sec:impact-model}

\subsection{Obtaining experimental data}
\label{sec:injecting}

\begin{figure*}
\begin{minipage}{0.5\textwidth}
\centering
\includegraphics[width=0.95\textwidth]{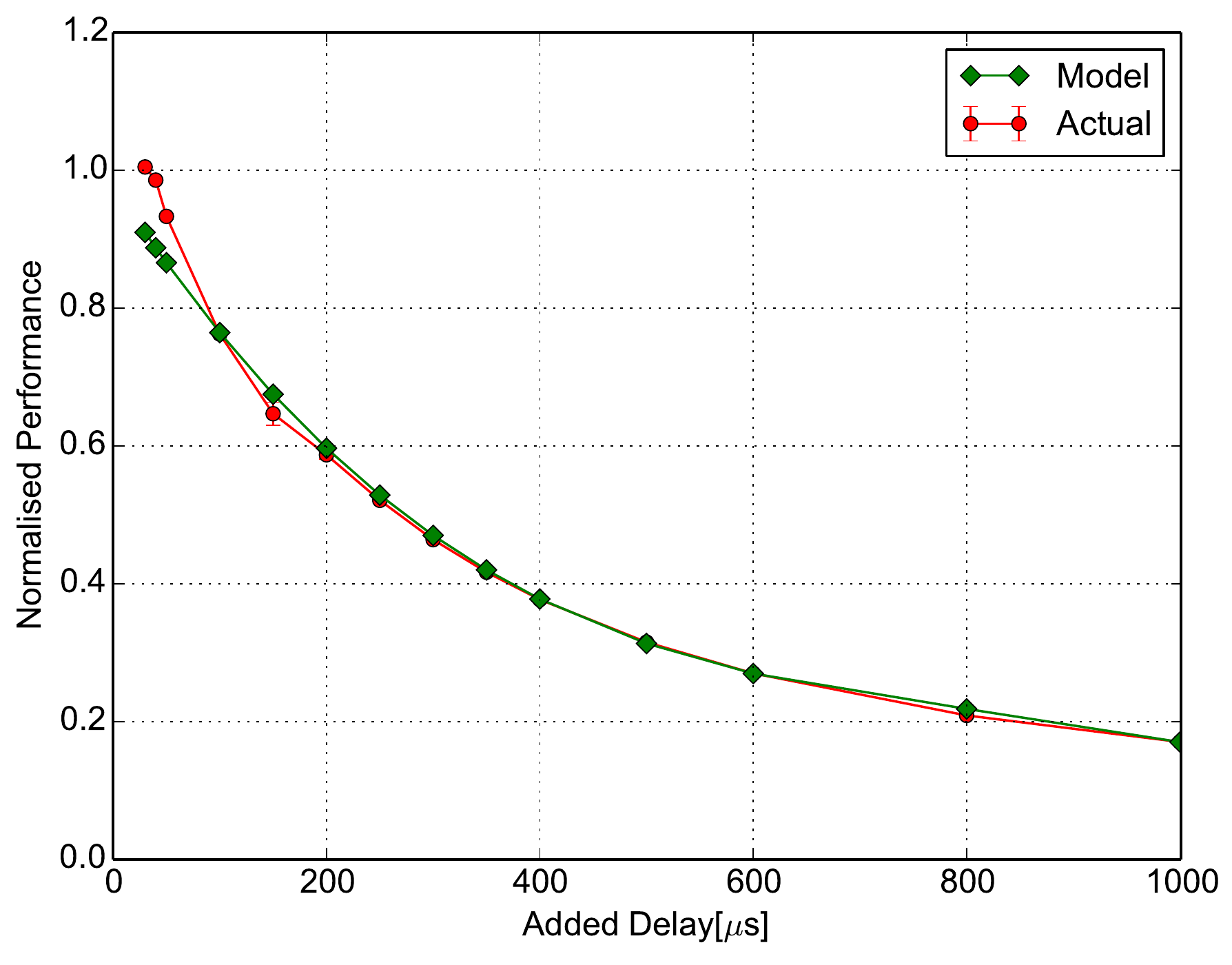} 
\subcaption{Memcached}
\label{fig:model-2d-memcached}
\end{minipage}
\begin{minipage}{0.5\textwidth}
\centering
\includegraphics[width=0.95\textwidth]{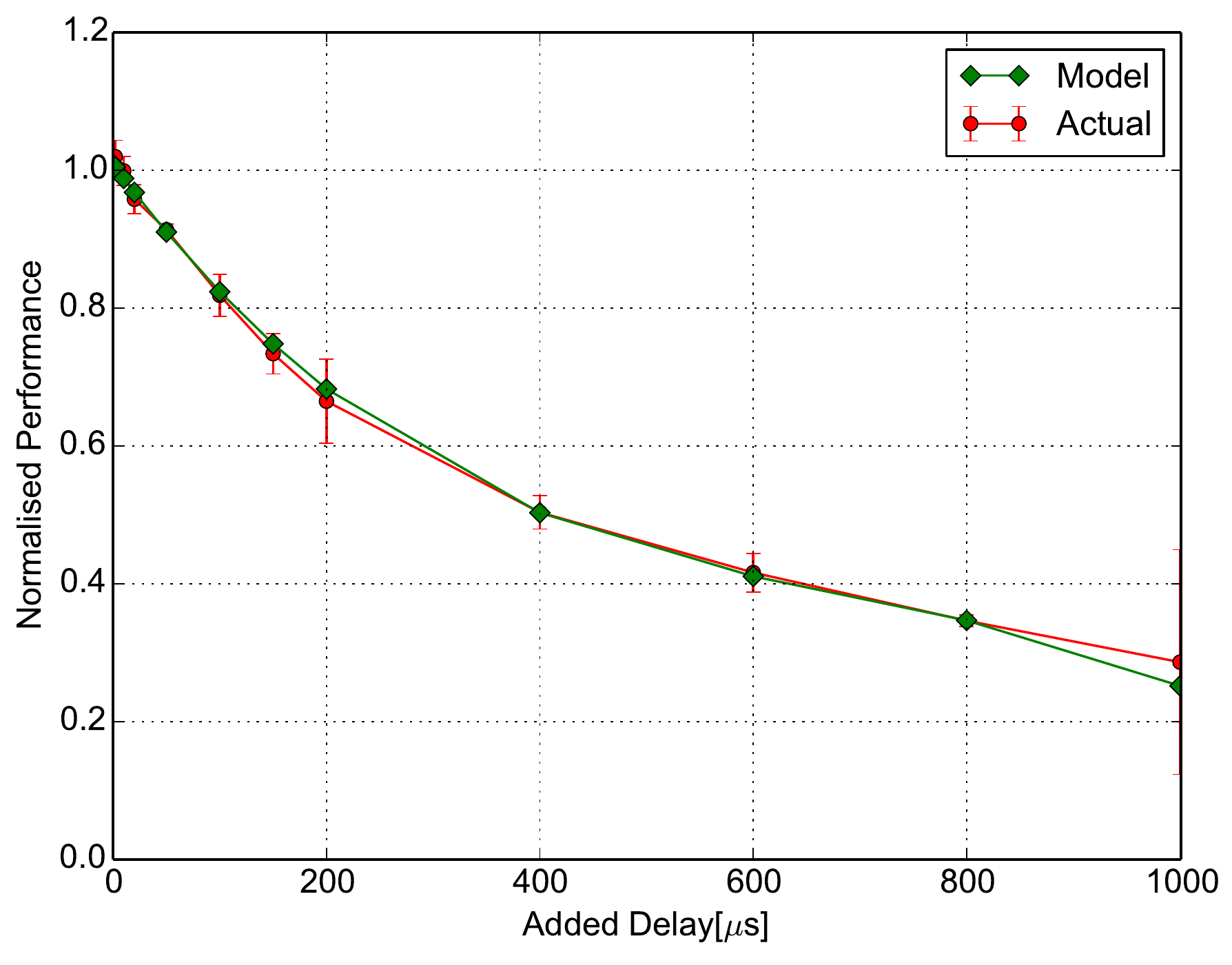} 
\subcaption{STRADS Lasso Regression}
\label{fig:model-2d-strads}
\end{minipage}
\begin{minipage}{0.5\textwidth}
\centering
\includegraphics[width=0.95\textwidth]{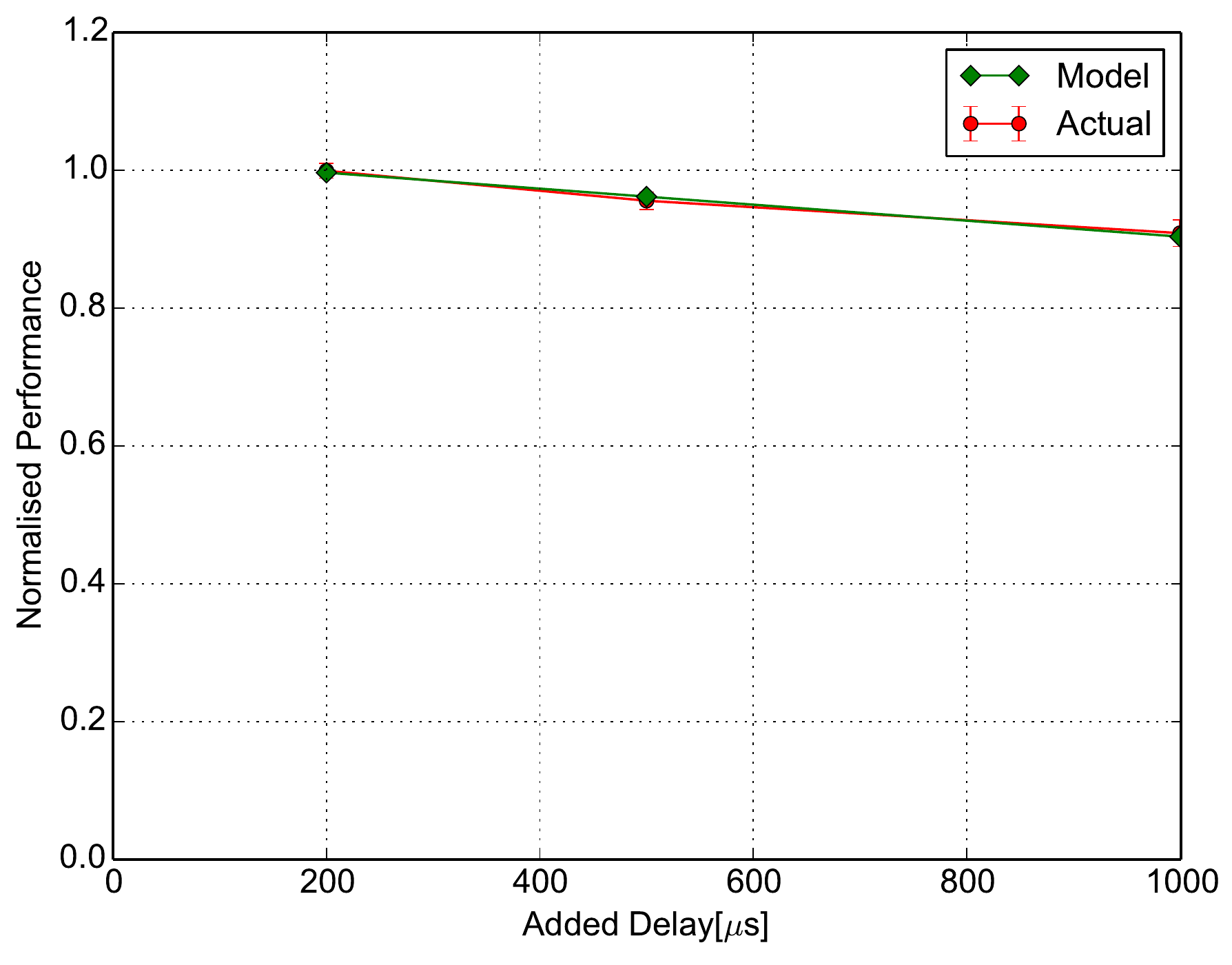} 
\subcaption{Spark GLM}
\label{fig:model-2d-spark}
\end{minipage}
\begin{minipage}{0.5\textwidth}
\centering
\includegraphics[width=0.95\textwidth]{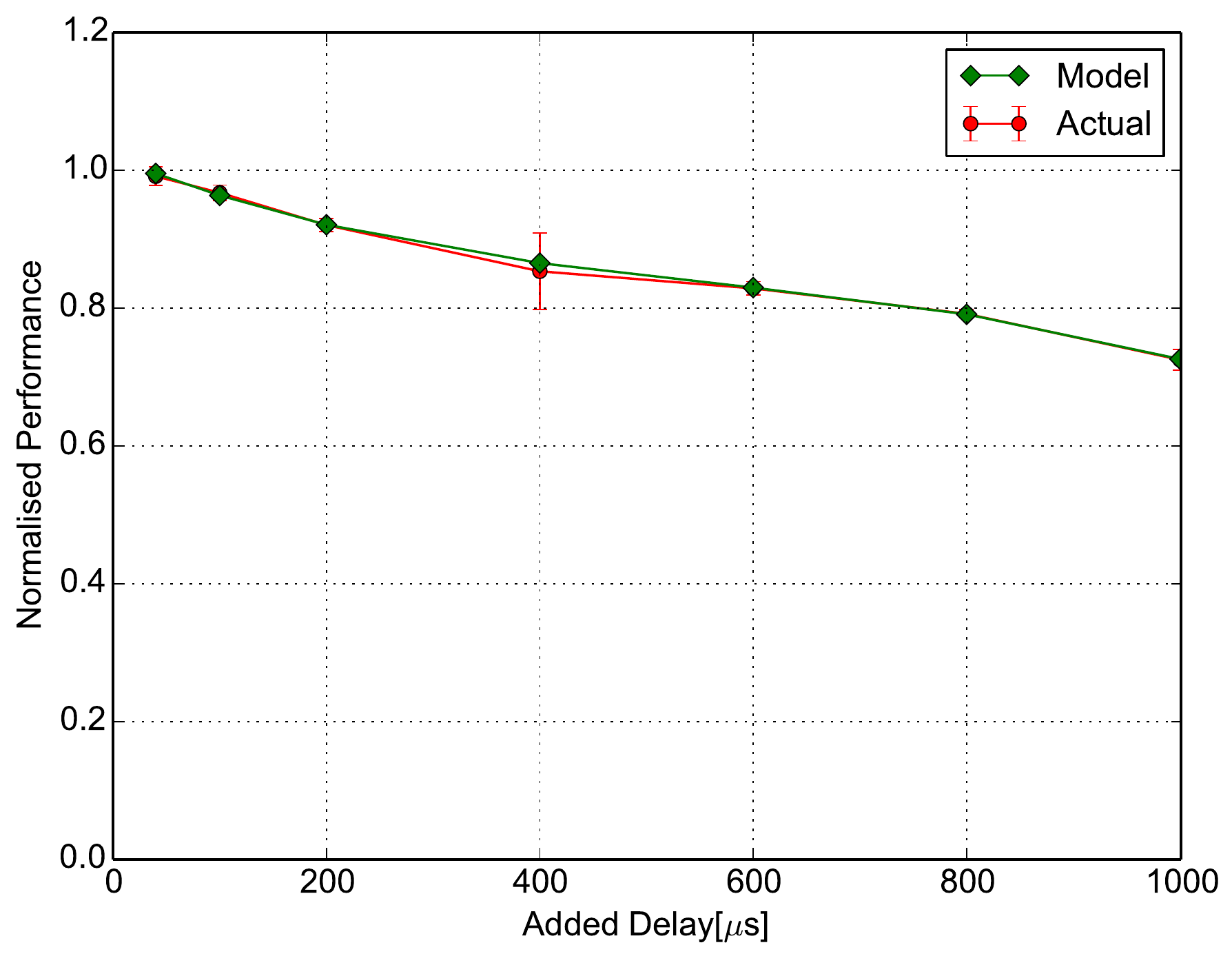} 
\subcaption{Tensorflow MNIST}
\label{fig:model-2d-tensorflow}
\end{minipage}
\caption{Applications experimental results (actual) and model on the results (model).}
\label{fig:all-results}
\end{figure*}

To determine the relationship between network latency and application performance, we used a methodology previously described in~\cite{pam2017, diana-report}, where we injected increasing amounts of latency in a networked system using a latency appliance~\cite{pam2017, latency-gadget, diana-report} that delays the packets by a fixed amount of time, and we measured how the application performance changes depending on the amount of inserted network latency. The experimental testbed is the one described in~\cite{diana-report}. Each host has an Intel Xeon E5-2430L v2 Ivy Bridge CPU with six cores, running at 2.4GHz with 64GB RAM, and is equipped with an Intel X520 NIC with two SFP+ ports, being connected at $10$Gbps through an Arista 7050Q switch. The hosts run Ubuntu Server 16.04, kernel version 4.4.0-75-generic.
Each application has a certain performance metric (see Table~\ref{tab:app-settings}). We first determined a baseline application performance, where the application runs on a number of hosts with optimal performance for the workload and on that specific setup (see Table~\ref{tab:app-settings}). After the baseline performance is determined for each application, we introduced a constant latency value between a host (on which the server/the master component of an application runs) and the other hosts (on which the clients/the workers of an application run) in both directions (send and receive), sweeping the range of values between $1\mu$s and $500\mu$s. 
Thus, the total latency values introduced range between $2\mu$s and $1000\mu$s.  
We chose values in this range based on the network latency values that we measured in different cloud providers in a previous work~\cite{diana-tma}. 
The application performance was measured for each injected latency value, and the experiments were run for each value a sufficient number of times to ensure reproducibility. The experimental results can be seen in Figure~\ref{fig:all-results} in the 'Actual' line. The complete results are presented in the technical report~\cite{diana-thesis}. 
This fixed amount of latency represents an amount of latency that can have different causes, such as physical distance, or the delay introduced by the network apparatus, \eg, NICs, switches.
However, increased network utilisation can lead to higher latency even if two hosts are close to each other. 

\begin{table*}
	\centering
        \small
	\setlength{\tabcolsep}{2pt}
	\begin{tabular}{|l||l|l|l|l|l|l|}
		\hline
		\textbf{Application}& \textbf{Host's role} & \textbf{\#Hosts} &\textbf{Performance} &\textbf{Runtime} & \textbf{Workload/} & \textbf{Dataset} \\ 
		& & & \textbf{Metric} & \textbf{Target} & \textbf{Dataset}& \textbf{Size}\\\hline
		\hline
     	Memcached~\cite{memcached}       & Server  & 5 & Queries/sec     & 10 seconds      & FB ETC ~\cite{Atikoglu:2012}    &     see~\cite{Atikoglu:2012} \\ \hline   	
        STRADS~\cite{Kim:2016}     & Coordinator      & 6 & Training time &  100K iterations   & Lasso Regression &  10K samples, 100K features \\ \hline
        Spark~\cite{spark} & Master & 8 & Training time &  100 iterations  & GLM Regression~\cite{spark-perf}   & 100K samples, 10K features   \\ \hline
        Tensorflow~\cite{tensorflow} & Master & 9 & Training time & 20K iterations & MNIST~\cite{mnist} & 60K examples\\ \hline
	\end{tabular}
	\caption{Applications settings.\#Hosts indicates the minimum number of hosts required to achieve optimal performance for the workload.}
	\label{tab:app-settings}
\end{table*}

\subsection{Modeling performance}

We now construct a \emph{function that predicts application performance dependent upon network latency} for each application. To model the relationship between network latency and application performance, we use SciPy's~\cite{scipy} \texttt{curve\_fit} function, which uses non-linear least squares to fit a function \emph{p}, to the experimental data. The \texttt{curve\_fit} returns optimal values for the parameters, so that the sum of the squared error of $\mathit{p}(x\_data, parameters) - y\_data$ is minimised. 
The function has one independent variable, the static latency, and the dependent variable is the
application's performance metric. We additionally use the standard deviation of the results as a parameter for the \texttt{curve\_fit} function.
We first normalise the values of the application performance with respect to the baseline performance. We then model the relationship between network latency and normalised application performance by fitting a polynomial function to the results, where the independent variable is the static latency, and the dependent variable is the normalised performance:
\begin{equation}
Normalised\_performance = \mathit{p}(static\_latency)
\end{equation}


\subsubsection{Key-value store: Memcached}
Memcached~\cite{memcached} is a widely used, in-memory, key-value store for arbitrary data. 
Clients can access the data stored in a memcached server remotely over the network. 
We use the Mutilate~\cite{mutilate} Memcached load generator, with the {\it Facebook ``ETC'' workload}, taken from~\cite{Atikoglu:2012}, which is considered representative of general-purpose key-value stores.
We consider Memcached's application performance metric the number of queries per second (QPS). 
The resulting model is shown in Figure~\ref{fig:model-2d-memcached} and in Equation~\ref{eq:memcached}.
This model does not capture the baseline performance, nor small static latency values. 
Therefore, the model needs to have two functions: a constant function, whose value is the baseline performance, and a logarithmic function fit on the experimental data. The first function gives the performance up to the threshold latency value beyond which the application performance starts to drop, \eg, $40$\micros. The Figure~\ref{fig:model-2d-memcached} does not show the first function.

\begin{equation}\label{eq:memcached}
\small
p(x) = \begin{cases}
		1, x < 40 \\
		1.067 - 3.093 \times 10^{-3} \times x + 4.084 \times 10^{-6} \times x^2 - \\ \; \; \; \; - 1.898 \times 10^{-9} \times x^3, x \geq 40
		\end{cases}
\end{equation}




\subsubsection{Machine Learning applications}

\paragraph{STRADS Lasso Regression}

STRADS~\cite{Kim:2016, strads} is a distributed framework for machine learning algorithms targeted to moderate cluster sizes between 1 and 100 machines.  
We evaluate the impact of network latency on the Lasso Regression~\cite{lasso} application implemented in this framework.
The network communication pattern can be represented as a star, with a central coordinator and scheduler on the master server, while workers communicate only with this master server. 
Since we do not allow the use of stale parameters during iterations (no pipelining) and the injected network latency does not change the scheduling of the parameters, the application performance metric can be represented by the job completion time, named in the next sections the \emph{training time}.  
The results are shown in Figure~\ref{fig:model-2d-strads} and in Equation~\ref{eq:strads}. The first function is the constant baseline performance up to $20$\micros.

\begin{equation}\label{eq:strads}
\small
p(x) = \begin{cases}
		1, x < 20\\
		1.009 - 2.095 \times 10^{-3} \times x + 2.571 \times 10^{-6} \times x^2 - \\ \; \; \; \; - 1.232 \times 10^{-9} \times x^3, x \geq 20
		\end{cases}
\end{equation}


 

\paragraph{Spark GLM Regression}

We use Apache Spark~\cite{spark}'s machine learning library (MLlib), on top of which we run the GLM regression benchmark from Spark-Perf~\cite{spark-perf}. We run Spark 1.6.3 in standalone mode.    
Spark follows a master-worker model. Spark supports broadcast and shuffle, which means that the workers do not communicate only with the master, but also between themselves, but even so, small latencies injected between every pair of hosts do not affect its performance.
We use as application performance metric the \emph{training time}, \eg, the time taken to train a model. The results are shown in Figure~\ref{fig:model-2d-spark} and in Equation~\ref{eq:spark}. The first function is the constant baseline performance up to $200$\micros.

\begin{equation}\label{eq:spark}
\small
p(x) = \begin{cases}
		1, x < 200\\
		-1.161 \times 10^{-4} \times x + 1.0199, x \geq 200
		\end{cases}
\end{equation}


\paragraph{Tensorflow Handwritten Digits Recognition}

Tensorflow~\cite{tensorflow} is a widely used machine learning framework. We use the MNIST dataset~\cite{mnist} for the handwriting recognition task as input data, and Softmax Regression for the training of the model. Tensorflow follows a master-worker model. The application performance metric used is the~\emph{training time}, similarly to Spark's performance metric. We use the \texttt{synchronise replicas} option, which means that the parameter updates from workers are aggregated before being applied in order to avoid stale gradients. 
The results are shown in Figure~\ref{fig:model-2d-tensorflow} and in Equation~\ref{eq:tensorflow}. The first function is the constant baseline performance up to $40$\micros.

\begin{equation}\label{eq:tensorflow}
\small
p(x) = \begin{cases}
		1, x < 40\\
		1.005 - 5.146 \times 10^{-4} \times x + 5.837 \times 10^{-7} \times x^2 -  \\ \; \; \; \; - 3.46 \times 10^{-10} \times x^3 
		, x \geq 40
		\end{cases}
\end{equation}


\subsubsection{Model generality}

Finding a general relationship between an application's performance and network latency is not easy. We sought to limit the influence of other factors (OS impact on application, number of cores, number of machines in the setup) on the measured application performance, leaving only the effect of network latency on the application performance. 
\begin{figure}
 \centering
        \includegraphics[width=0.5\textwidth]{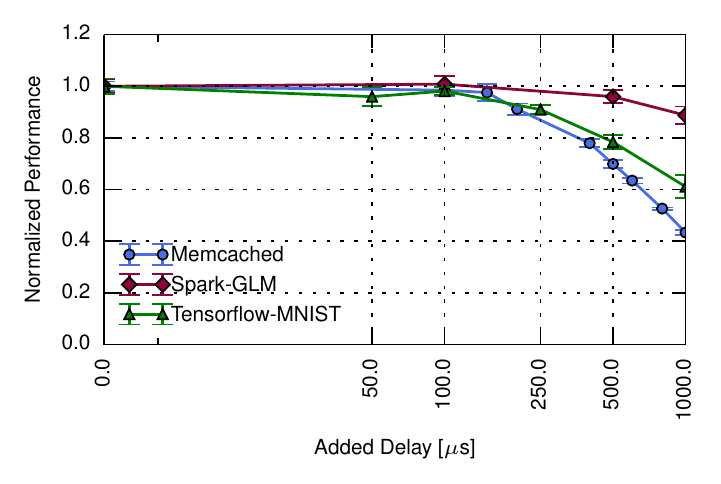}
        \caption{The effect of injected static latency on typical cloud applications' performance running on cloud hardware.}
       \label{fig:cloud-netem-results}
\end{figure}

While the impact of latency on performance evaluated in Section~\ref{sec:injecting} was conducted on a specific setup, the results offer an intuition on the application behaviour that can be generalised to other setups and scenarios as well. 
Even though there are differences between computing platforms, the results have the same scale and follow the same trends. We exemplify this statement by evaluating three of the selected applications (Memcached, Spark GLM Regression and Tensorflow MNIST) on a different setup in a \dc in Microsoft Azure.
The setup has one server/master VM and five clients/workers VMs. The VM type is Standard E16s v3 with 16 virtual CPUs and 128 GB memory. We insert latencies of over $100$\micros with NetEm~\cite{netem} at every host. 
The general trend in Figure~\ref{fig:cloud-netem-results} is the same as in Figure~\ref{fig:all-results} for the selected applications. In the case of Spark GLM Regression and Tensorflow MNIST, the drop in performance on this setup is steeper than on the local testbed. On the other hand, the Memcached server is less affected on this setup compared with the local testbed. These differences can be the result of any or all of the following factors: virtualisation, different number of hosts, host specifications, different network topology, varying network utilisation due to the shared network in the cloud, latency injection through software emulation instead of through a hardware-based solution. 

By fitting a naive model to the experimental datasets determined in this section, we took a first step towards building a more general model.
A system that benchmarks the application performance under different configurations and network conditions would help in building a complete application performance profile. Such a system would be used to predict the application performance under different circumstances. The predictions can then be used to inform cloud operators and users in order to find the optimal placement under given network conditions.




\section{Background on cluster scheduling}
\label{sec:firmament-background}

\begin{figure*}
\centering
\includegraphics[width=0.8\linewidth]{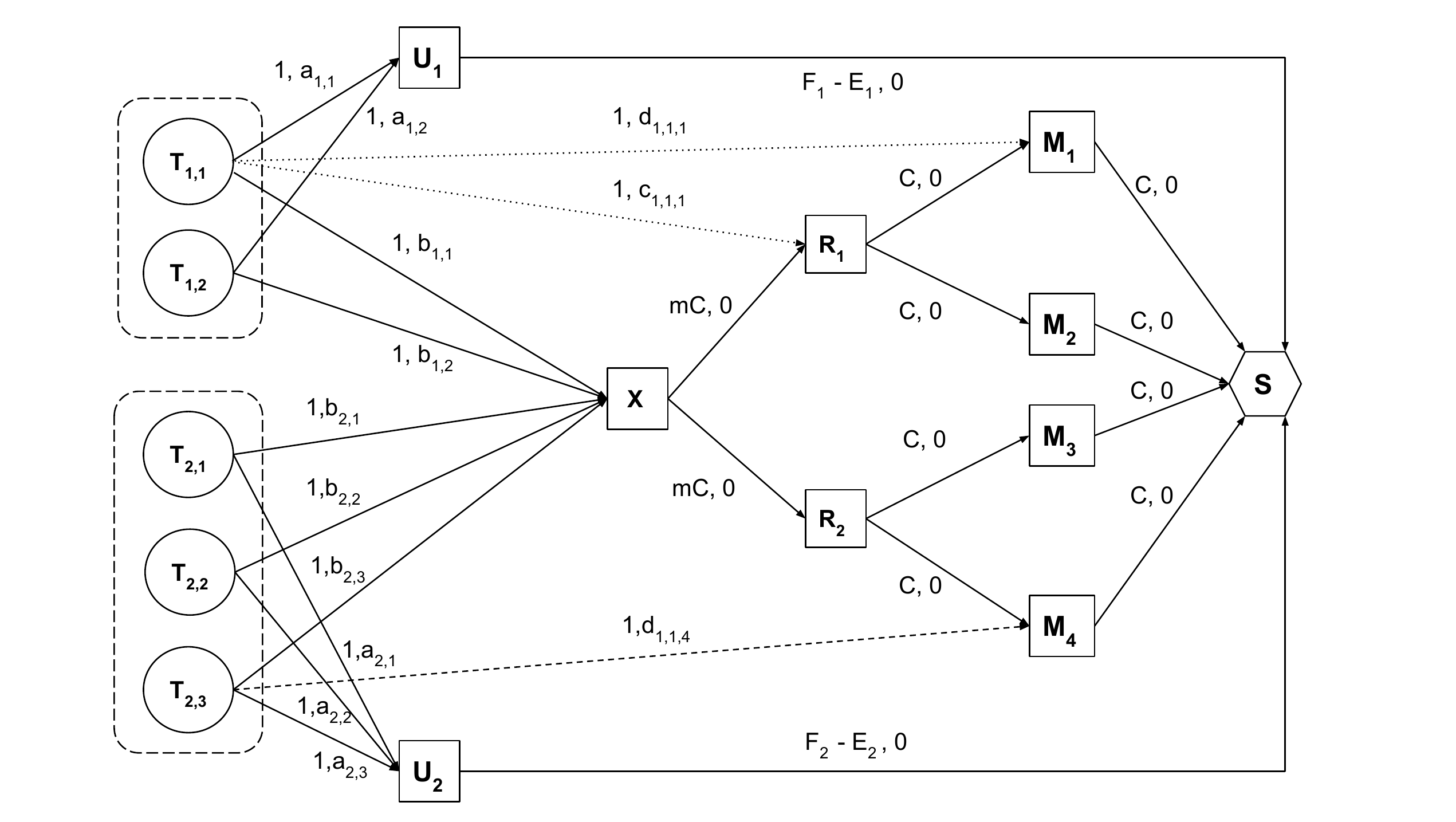}
\caption{A general flow network with annotated capacities and costs on arcs. Job $J_{1}$ has tasks $T_{1,1}$ and $T_{1,2}$. Job $J_{2}$ has tasks $T_{2,1}$, $T_{2,2}$ and $T_{2,3}$. The unscheduled aggregator is $U_{1}$. The machines in the cluster are $M_{1}$, $M_{2}$, $M_{3}$ and $M_{4}$. Rack aggregators are $R_{1}$ and $R_{2}$. The cluster aggregator is $X$. The sink vertex is $S$.}
\label{fig:general-task-graph}
\end{figure*}

NoMora extends the Firmament cluster scheduler~\cite{firmament}. We chose to extend Firmament because it is a centralised scheduler that considers the entire workload across the whole cluster, making it straightforward to incorporate the network latency measured between every pair of hosts in the cluster, and due to its low latency (sub-second) task placement~\cite{firmament}. Firmament exposes an API to implement scheduling policies, that may incorporate different task constraints. A scheduling policy defines a flow network representing the cluster, where the nodes define tasks and resources. The policy can also use task profiles to guide the task placement through preference arcs to machines that meet the criteria desired by the task. Events such as \emph{task arrival}, \emph{task completion}, \emph{machine addition to the cluster}, or \emph{machine removal from the cluster}, change the flow network.
When cluster events change the flow network, Firmament's min-cost flow solver computes the optimal flow on the updated flow network. 
The updates to the flow network caused by the cluster events are not applied while the solver runs, but only after the solver finishes computing the optimal solution. 
After the solver finishes running, Firmament extracts the task placements from the optimal flow, and applies these changes in the cluster. We next give an overview of how the cluster scheduling problem is mapped to the minimum-cost maximum-flow optimisation problem, as described in Quincy~\cite{quincy} and Firmament~\cite{firmament}.

\textbf{Flow network} 
\label{ss:flow-network}
Firstly, we provide a high-level overview of the structure of the flow network, which can be seen in Figure~\ref{fig:general-task-graph}. By \emph{flow network} we refer to a directed graph where each arc has a capacity and a cost to send flow across that arc.
Each submitted task $T_{i,j}$, representing task $j$ of job $J_{i}$, is represented by a vertex in the graph, and it generates one unit of flow. The sink $S$ drains the flow generated by the submitted tasks. A task vertex needs to send a unit of flow  along a path composed of directed arcs in the graph to the sink $S$. The path can pass through a vertex that corresponds to a machine (host) $M_{m}$, meaning the task is scheduled to run on that machine, or it can pass through a special vertex for the unscheduled tasks of that job $U_{i}$, meaning that the task is not scheduled. In this way, even if the task is not scheduled to run, the flow generated by this task is routed through the unscheduled aggregator to the sink.

The graph can have an arc between every task and every machine, but this would make prohibitive the computation of an optimal scheduling solution in a short time, as the graph would scale linearly with the number of machines in the cluster. To reduce the number of arcs in the graph, a cluster aggregator $X$ and rack aggregators $R_{r}$ have been introduced in Quincy, inspired by the topology of a typical \dc. The cost of the arc between a task and the cluster aggregator is the maximum cost across all of the machines in the cluster. Similarly, the cost of the arc between a task and a rack aggregator $R_{r}$ is the maximum cost across all of the machines in that rack. It can be easily seen that the costs to the cluster and rack aggregators serve as a conservative approximation, providing an upper bound for a set of resources that are grouped together. 

\textbf{Capacity assignment} Each arc in the flow network has a capacity $c$ for flow, bounded by $c_{min}$ and $c_{max}$. In Firmament and Quincy, $c_{min}$ is usually zero, while $c_{max}$ depends on the type of vertices connected by the arc and on the cost model. Given that the minimum capacity is zero, it is omitted from here onwards, with only the maximum capacity values being presented. The capacity of an arc between a task and any other vertex is $1$. If a machine has $C$ cores and a rack has $m$ machines, the capacity of an arc between a rack aggregator and a machine is $C$, and the capacity of an arc between the cluster aggregator and a rack is $C\times m=Cm$. 
The capacity of an arc between a machine and the sink is $C$. The capacity between an unscheduled aggregator $U_{i}$ and the sink $S$ is represented by the difference between the maximum number of tasks to run concurrently for a job $J_{i}$, $F_{i}$, and the minimum number of tasks to run concurrently for job $J_{i}$, $E_{i}$, with $0 \leq E_{i} \leq F_{i} \leq N_{i}$, where $N_{i}$ is the total number of tasks in job $J_{i}$. These limits can be used to ensure a fair allocation of runnable tasks between jobs~\cite{quincy}. In the NoMora policy, this capacity is set to $1$.

\textbf{Cost assignment} The cost on an arc represents how much it costs to schedule any task that can send flow on this arc on any machine that is reachable via this arc. 

\emph{Task to machine arc:} The cost on the arc between a task vertex $T_{i,j}$ and a machine vertex $M_{m}$ is denoted by $d_{i,j,m}$, and is computed according to information regarding the task and machine. In most cases, this cost is being decreased by 
how much the task has already run, $\beta_{i, j}$ . 

\emph{Task to resource aggregator arc:} The cost on the arc between a task vertex $T_{i,j}$ and a rack aggregator vertex $R_{r}$, denoted $c_{i,j,r}$, represents the cost to schedule the task on any machine within the rack, and is set to the worst case cost amongst all costs across that rack. The cost on the arc between a task vertex $T_{i,j}$ and the cluster aggregator $X$, denoted by $b_{i,j}$, represents the cost to schedule the task on any machine within the cluster, and is set to the worst case cost amongst all costs across the cluster. 

\emph{Task to unscheduled aggregator arc:} The cost on the arc between a task vertex $T_{i,j}$ and the unscheduled aggregator $U_{i}$, denoted by $a_{i,j}$, is usually larger than any other costs in the flow network. The cost on this arc increases as a function of the task's wait time, in order to force the task to be scheduled, and it is scaled by a constant \emph{wait time factor} $\omega$, which increases the cost of tasks being unscheduled. 

\emph{Preemption:} If preemption is enabled, the scheduler can preempt a task that it is running on a machine, which means the flow pertaining to that task is routed via the unscheduled aggregator, or migrate the task to a different machine, meaning that the flow is routed via that new machine's vertex. If preemption is not enabled, then a scheduled task will have in the flow network only the arc to the machine that it is currently running on, with all the other arcs being removed once the task is scheduled.

\section{NoMora}
\label{sec:nomora}

\subsection{Architecture}

We combine the following three elements in the NoMora cluster scheduling architecture, where (1) and (2) offer inputs to (3): (1)
functions that predict application performance dependent upon network latency; (2) network latency measurement system; (3) the latency-driven application performance-aware cluster scheduling policy implemented on top of the Firmament cluster scheduler. 
The functions that predict application performance dependent upon network latency were determined in Section~\ref{sec:impact-model}. 
The second component of the architecture is the network latency measurement system.
Systems such as PTPmesh~\cite{diana-mascots}, Pingmesh~\cite{pingmesh} or NetNORAD~\cite{netnorad}, can provide the most recently measured network latency between hosts. Due to the scale of \dcs, these systems do not run probes between every pair of hosts, but instead choose fewer hosts to ensure the largest coverage. Additionally, they set a minimum probing interval to bound the network traffic generated. The data collected by these systems is fed to the cluster scheduler to aid in making task placement or migration decisions.
Default latency values can be determined based on the network topology of the \dc to be used instead of the actual measured latencies if these are unavailable.
The third component of the system, the policy, is discussed in the following section.

\subsection{Latency-driven, application performance-aware, policy}

We propose a new \emph{latency-driven, application performance-aware, policy} whose goal is to place distributed applications in a \dc in a manner that gives them improved application performance. This generally leads to grouping tasks as close as possible, in a rack or on the same machine, for the applications for which latency matters, such as Memcached or machine learning frameworks (STRADS, Tensorflow). For the tasks that do not fit within the same rack or on the same machine, the policy finds the machine that offers the best application performance amongst the available placements. On the other hand, applications like Spark for which additional latency of up to one millisecond does not matter will have more freedom when being placed within the \dc. Furthermore, if the network conditions change, a task whose performance degrades can be migrated to a better placement. 

Since the applications we studied in Section~\ref{sec:impact-model} are client-server applications or worker-master applications, 
we consider that the server/the master has a special role, because it has to be running before the clients/workers. 
We call the server (for client-server applications) / the master (for master-workers applications) the \emph{root task}. 
Thus, the policy needs to schedule the root task first. The root task is scheduled immediately in any place available in the cluster.
The other tasks of the job (clients/workers) are not scheduled until the root task is scheduled. While this adds delay in scheduling for these tasks, the delay is minimal, since they will be scheduled in the next scheduling round based on the placement of the root task. 
In our policy, a task's placement does not depend on a machine's architectural properties (CPU, RAM, etc.) or on the properties of the tasks that run a machine, but on the placement of another task (the root task), and on the network latency between a machine considered for the task's placement and the root task's placement. 
Our policy uses to assign costs to arcs the application performance predictions and network latency measurements between hosts to determine the expected application performance. 

In summary, the placement of a task follows these steps:
\begin{enumerate}
	\item the root task is scheduled on any available machine;
	\item if a task that is not a root task enters the system at the same time as the root task, or before the root task is scheduled, it will not be scheduled, waiting instead;
	\item if the root task is scheduled, then a new task's placement is determined based on the application performance prediction, and current network latencies to the root task's placement.
\end{enumerate}

\textbf{Flow network} The flow network is similar to the one in Figure~\ref{fig:general-task-graph}. Arcs are defined between a task and the potential machines on which it can run, and each arc has a cost computed using the application performance predictions dependent upon network latency for each application. 
When a job is submitted, the root task $T_{i,0}$ is assigned a single arc to the cluster aggregator, with a cost of $0$, which means that the root task will be scheduled immediately on any available machine. After it is scheduled, the root task will have an arc from the root task to the machine it is running on. The other tasks of the job will wait for the root to be scheduled first, and they do not have any arcs initially. After the root task is scheduled, each task $T_{i, j}$ will have preference arcs to the cluster aggregator $X$, to rack aggregators $R_{r}$ and machines $M_{m}$ based on the cost to schedule the task on those resources and on the parameters of the policy.

\textbf{Cost assignment} The cost assignment is also called \emph{cost model}~\cite{firmament}. NoMora's cost assignment is similar to Quincy's cost assignment~\cite{quincy}. The Quincy policy considers in the cost computation data locality, task wait time (the time a task waits before being scheduled) and how much time a task has run before being preempted (\emph{preemption cost}). These factors are considered because having good data locality reduces the job's runtime in the scenario considered by Quincy, but finding a good placement can mean waiting more time until a suitable machine is free (the task wait time increases), or preempting a task that is already running (if this task is restarted from the beginning on another machine, then the time the task has already run is lost). Instead of considering data locality, in NoMora we consider the cost to the root task computed based on application performance prediction dependent upon network latency for a job (as built in Section~\ref{sec:impact-model}), combined with measured network latency between the root task machine and the machine under consideration. 
Similarly to the Quincy cost model, in NoMora we also factor the task wait time when computing the cost of the arc to the unscheduled aggregator, and, if preemption is enabled, the preemption cost in the arc cost computation. 

\begin{table}
\begin{center}
\begin{tabular}{|c||c|c|}
\hline 
\textbf{Arc} & \textbf{Capacity} & \textbf{Value} \\
\hline
\hline
$T_{i,j}\rightarrow U_{i}$ & $1$ & $a_{i,j}$ \\
\hline
$T_{i,j}\rightarrow X$ & $1$ & $b_{i,j}$ \\
\hline
$T_{i,j}\rightarrow R_{r}$ & $1$ & $c_{i,j,r}$ \\
\hline
$T_{i,j}\rightarrow M_{m}$ & $1$ & $d_{i,j,m}$ \\
\hline
$X\rightarrow R_{r}$ & $mC$ & $0$  \\
\hline
$R_{r}\rightarrow M_{m}$ & $C$ & $0$ \\
\hline
$M_{m}\rightarrow S$ & $C$ & $0$ \\
\hline
$U_{i}\rightarrow S$ & $1$ & $0$ \\
\hline
\end{tabular}
\end{center}
\caption{Arcs in the NoMora flow network.}
\label{tab:nomora-costs}
\end{table}

Assuming the root task is running on machine $M_{root}$ and a task $j$, $T_{i, j}$, of job $J_{i}$ can be scheduled on machine $M_{m}$, then the cost of the arc from $T_{i, j}$ to $M_{m}$ is:

\begin{equation}
d_{i,j,m} = cost(T_{i, j}, M_{m}) = \frac{1}{p(max(latency(M_{root}, M_{m})))}
\end{equation}

where $p(max(latency(M_{root}, M_{m})))$ is the expected application performance for the measured network latency between machine $M_{root}$ and machine $M_{m}$, as determined in Section~\ref{sec:impact-model}. I invert the performance because, when the performance is smaller, the cost assigned to the arc is higher, making the machine to which the arc points to less desirable for running the task on. Since in \dcs typically there are multiple paths between two machines, in order to be conservative, we use the maximum latency value measured between the two machines because, due to ECMP, we cannot know which of the available paths the application's flows will take. 

Preempting a task presents a trade-off between migrating the task to a better placement and the amount of time the task has already run (on the current machine or on a different one)~\cite{vmm}. If preemption is enabled, the amount of time the task has already run, $\beta_{i, j}$, can be subtracted from the $cost(T_{i, j}, R_{r})$. This leads to less task migrations happening, because it becomes less advantageous to preempt a task and restart it on another machine after migration as more time the task is running, essentially wasting the work that has already been done. 

\begin{equation}
d_{i,j,r} = cost(T_{i, j}, M_{m}) - \beta_{i, j}
\end{equation}

Similarly, the cost of the arc from $T_{i, j}$ to rack $R_{r}$ is the cost to the worst cost machine in rack $r$:

\begin{equation}
c_{i,j,r} = cost(T_{i, j}, R_{r}) = \max\limits_{m \in r} \frac{1}{p(max(latency(M_{root}, M_{m})))}
\end{equation}

where $p(max(latency(M_{root}, M_{m})))$ is the expected application performance for the measured network latency between machine $M_{root}$ and a machine $M_{m}$ in rack $r$. Similarly, to be conservative due to ECMP, we take the maximum value of the latencies between $M_{root}$ and $M_{m}$. 

The cost to the cluster aggregator $X$ is the cost to the worst cost rack, which is obtained by taking the maximum of the costs to racks.

\begin{equation}
b_{i,j} = \max\limits_{r} c_{i, j, r}
\end{equation}

The cost to the unscheduled aggregator $U_{i}$ is computed using the task's wait time, $\alpha_{i, j}$, scaled by a constant factor $\omega$~(\cref{sec:firmament-background}), to which a constant cost factor $\gamma$, that is larger than any other possible arc costs, is added.   

\begin{equation}
a_{i,j} = \omega \times \alpha_{i, j} + \gamma
\end{equation}

The costs on the arcs are rounded to two significant digits, and then multiplied by a factor of $100$, since the costs must be integer numbers for the solver to understand. For a performance of $1$, the cost is $\frac{1}{1} \times 100 = 100$.  For a performance of $0.1$, the cost is $ \frac{1}{0.1} \times 100 = 1000$. The cost to the unscheduled aggregator is offset by $\gamma$, greater than all the other possible costs. 


Since the network latency is not constant in a \dc, as shown in Section~\ref{sec:motivation}, the costs associated with the arcs are updated based on the latest measured network latency values, and as a result, the preference arcs for the tasks are updated. If preemption is enabled, the cost of the arcs for a running task will also be updated.	

\textbf{Cost model parameters}
\label{sec:cost-model-params}
The cost model has two main parameters: $p_{m}$, threshold for the cost on an arc to a machine in order for that machine to be on the
preferred list of machines on which the task can run, and $p_{r}$,
threshold for the cost on an arc to a rack in order for that rack to be on the preferred list
of racks in which the task can run.
The first preference list comprises the machines on which the application may run to achieve the desired performance. This list should be kept small for the scheduling latency to take a reasonable amount of time. But having a small preference list means the application's placement options are limited. To mitigate this, the second preference list, which comprises the racks on which the application may run, was introduced. 
The second list is smaller than the first one, since the number of racks is smaller than the number of machines. This allows a bigger threshold to be set for the second parameter of the model, offering more placement options for the application's tasks, while keeping the first preference list small. 

\section{NoMora evaluation}
\label{sec:nomora-eval}

We evaluate NoMora in simulation, using the same simulator as for Firmament's evaluation, extended to provide network latency measurements between pairs of hosts, and to update them during the simulation. Secondly, we added application performance predictions dependent upon network latency per job and per task (same function for all tasks of a job). Finally, we implemented the policy that uses these predictions and the latency measurements to compute task placements.

\textbf{Cluster workloads} 
No information about the network communication patterns between a job's tasks, nor about their sensitivity to network latency are provided in public cluster workloads. Thus, we have assigned the network latency to application performance functions determined in Section~\ref{sec:impact-model} to the jobs in the Google workload~\cite{google-workload}. We did not include the single task jobs, as they do not communicate with any other task. We used 24 hours of the trace.

\textbf{Application performance predictions dependent upon network latency} The predictions are discretised in steps of 10\micros, and are stored in a hash table for each job. The network latency value between two machines is rounded to the nearest latency value for which the prediction function has an entry in the hash table. For the latency values in the used traces that are outside the interval of defined values, we use the smallest performance value defined for that function.
The different prediction functions are assigned randomly in different proportions to the jobs. For the experiments presented in this section, $50$\% of the jobs use the Memcached prediction, $25$\% of the jobs use the STRADS prediction and $25$\% the Tensorflow prediction. This scenario is one of the most challenging, as Memcached is the most latency sensitive application that we studied. we did not use the Spark prediction, which is almost constant, as it would not be challenging to place such jobs. Given the functions built in Section~\ref{sec:impact-model}, for which the normalised performance
does not drop below 0.1, we set $\gamma=1001$ for the simulation. 

\textbf{Network latency measurements} The simulator leverages the network latency measurements dataset from~\cite{diana-tma}. With $18$ week-long traces provided, we further divide each trace in $7$ (for each day of the week), and we assign them to machine pairs considering the physical distance between servers as a criterion. Assuming a typical fat-tree topology for a \dc~\cite{fattree} and based on the latency values measured in Azure by~\cite{vnet-pingmesh}, we use the traces with the lowest values for machine pairs located in the same rack ($6$ traces - GCE), the traces with intermediate values for machine pairs located within the same pod ($6$ traces - Azure), and the traces with the largest values for machines located in different pods ($6$ traces - EC2). These traces are used to provide the latency values between hosts for the duration of the simulation, which is one day. Since we do not have different traces for each machine pair, we scale the values of each trace using a coefficient between 0.8 and 1.2, selected randomly for intra-pod and inter-pod values. For the traces within the rack, we scale them between 0.5 and 1. For the latency values between cores on the same server we use a small constant. Latency values from traces are provided every second in the simulation, in total $86,400$ per day between every pair of hosts in the cluster.

\textbf{Topology} We use the Google workload of 12,500 machines~\cite{google-workload}. The machines used in the workload are grouped into racks and pods at the beginning and during the simulation.
We set the number of machines per rack to 48, and the number of racks per pod to 16. 
The results will be influenced by the number of hosts per rack and the number of racks per pod due to the assignment of the different cloud latency traces. These two numbers were chosen to reproduce a small cluster. However, if there are more hosts per rack and more racks per pod, then there will be a greater chance to fit all the tasks of a job in the same rack or in the same pod, meaning the job will have a good overall application performance.

We performed experiments with the settings of the Facebook \dc topology (192 hosts per rack and 48 racks per pod)~\cite{facebook-dc}, but for a cluster of only 12,500 machines as the one presented in the Google trace, it means there is only one complete pod and an incomplete one, with a total of approximately 260 racks. The overall application performance in this case is very high due to the small network latencies assigned between the hosts within the same rack.

\textbf{Evaluation metrics}
Through the evaluation of the NoMora cluster scheduling policy, we seek to answer the following questions:
\begin{itemize}
\item does NoMora's placement improve application performance compared to a random placement policy and a load-spreading policy?
\item how long does it take to compute a placement solution?
\item how long does a task have to wait before it runs?
\end{itemize}

In order to know if our policy improves application performance compared to other policies, we compute the \textbf{average application performance}. This metric measures NoMora's task placement quality. It is computed as the application performance determined by the network latency in every measurement interval divided by the maximum application performance that could be achieved in every measurement interval, and it is computed for the job's total runtime. 

The \textbf{algorithm runtime} is the time it takes for Firmament's min-cost max-flow algorithm to run. We compare our policy's runtime with other Firmament policies' runtimes to ensure that our policy is scalable when run by a centralised cluster scheduler. Additionally, the time it takes to compute the applications' placements in a round gives an indication of the time interval that is needed between latency measurements. For example, if the algorithm runtime would be in the order of minutes, running latency measurements every few seconds would not be useful, since the measurements accumulated over the scheduler's runtime would not be used by the scheduler. On the other hand, if the algorithm runtime is in the order of milliseconds, running latency measurements every second or few seconds is useful, the scheduler being able to use the most recent measurement data.

The \textbf{task placement latency} is the time between task submission and task placement. This metric includes the task wait time, which should be as short as possible, but also the time it takes for Firmament to update the flow network, Flowlessly's runtime and the time to iterate over the placements computed by Flowlessly. In the context of our policy, the metric also captures by how long the tasks are delayed when they are waiting for their root task to be placed first before them. Another metric that we looked at is the \emph{task response time}, which is the time between a task's submission and its completion. 

\subsection{Placement quality}

We compare the NoMora policy, using different parameter values for the cost model (\cref{sec:cost-model-params}), with a random policy that uses fixed costs (tasks always schedule if resources are idle), and a load-spreading policy that balances the tasks across machines. We enable preemption only for the NoMora policy, since the two other policies would not benefit from preemption due to their different scheduling goals. The random policy schedules tasks if resources are available, thus migrating a task does not make sense, since the task is already running. The load-spreading policy schedules tasks based on current task counts on machines, thus potential task load imbalance can be handled by scheduling new tasks on less loaded machines instead of migrating already running tasks on the less loaded machines.  

\begin{figure}
 \centering
        \includegraphics[width=0.4\textwidth]{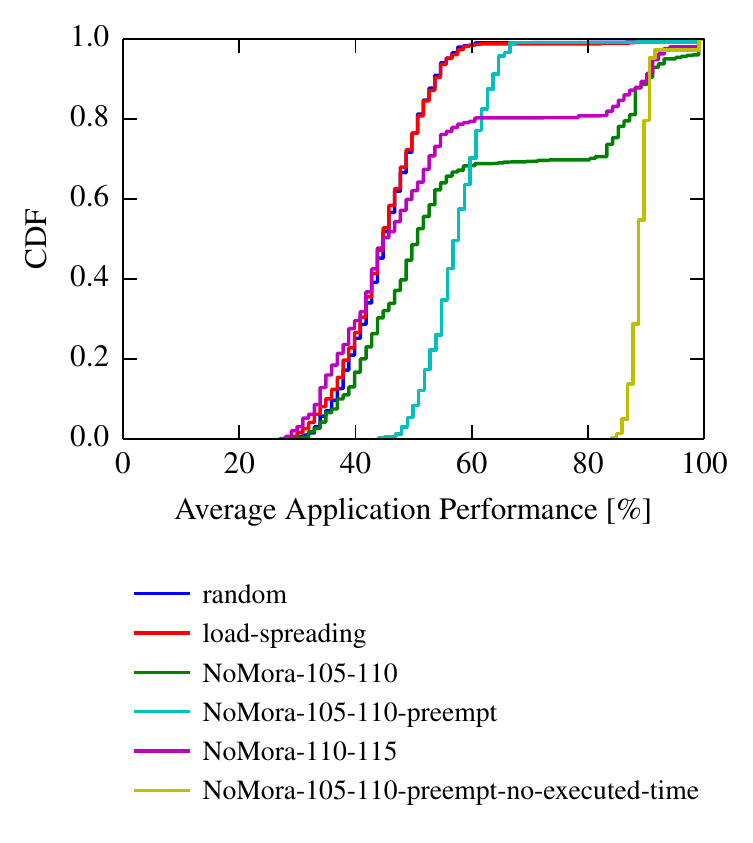}
        \caption{Average application performance for different policies on the Google workload.}
       \label{fig:google-placement-quality}
\end{figure}

The results for the average application performance for different policies can be seen in Figure~\ref{fig:google-placement-quality}.
we compute the area marked by the $y$-axis, the CDF, and the straight horizontal line with $y=1$, for each policy. According to this computation, the maximum area corresponding to the maximum average application performance across applications is 100\%, and it would be obtained for a vertical line at $x=100\%$. Next, we subtract from the NoMora policies areas the random and load-spreading areas to assess the placement improvement given by the NoMora policy. 
The total area for the random policy is 47.2\%, while for the load-spreading policy it is 46.8\%. For the NoMora policy with parameters $p_{m}=105$ and $p_{r}=110$ the area is 60.2\%, NoMora with parameters $p_{m}=105$ and $p_{r}=110$ and preemption enabled is 59\%, NoMora with parameters $p_{m}=110$ and $p_{r}=115$ is 51.85\%, and finally, NoMora with parameters $p_{m}=105$ and $p_{r}=110$, and preemption enabled with $\beta_{i, j}=0$, is $89.6$\%. The maximum overall improvement without preemption enabled is $13$\% over the random policy and $13.4$\% over the load-spreading policy, and is obtained for NoMora with parameters $p_{m}=105$ and $p_{r}=110$. If preemption is enabled and $\beta_{i, j}=0$ (the time already executed by a task is not considered in the arc cost computation), the improvement is considerable, $42.4$\% over the random policy, and $42.8$\% over the load-spreading one. 

The improvement in average application performance is not substantial when preemption is not enabled because of the root task's random placement, and also because of a smaller number of available places a task can be scheduled because of long-running jobs that are set up at the beginning of the trace. The tasks of the jobs are placed in the best available slots in relation to the root task's placement. In this way, we constrain the available placements, and the policy searches for placements in relation to a known location rather than trying to find the a placement for all the tasks of a job simultaneously. We further explain the reason behind this design decision and its implications in Section~\ref{sec:limitations}.

It can be seen that the CDF of NoMora with \emph{preemption enabled} has a different shape than the other policies. This is due to task preemption, which can correct the initial placement if it is not good (because of the random placement of the root task), and it can also migrate tasks when their current placement is not good anymore. The improvement provided by the NoMora policies without preemption is evident from Figure~\ref{fig:google-placement-quality}, but it can also be seen that the CDFs start at approximately the same value ($27$\%-$28$\% average application performance), and have an initially similar shape to the random and load-spreading CDFs. On the other hand, for NoMora with preemption enabled, the minimum average application performance is $44$\% and $84$\% respectively, which means that the improvement in application performance happens across all jobs due to migration to better placements.

\subsection{Algorithm runtime}

The algorithm runtime depends on the number of arcs from each task to the resources, but also on the cluster size. As the number of arcs or the cluster size increases, so does the algorithm runtime.
The two parameters of the cost model (\cref{sec:cost-model-params}) influence the number of arcs the graph has between task nodes and machine nodes or rack nodes, and hence the algorithm runtime, which depends on the flow network size and on the number of tasks considered per scheduling round. If the thresholds are lower, the preference lists will be smaller. In this case, the applications' performance will be higher (only high-quality placements considered), but they will have less placement options to be scheduled, and thus the wait time may increase. The tasks will have to wait for the machines that offer the performance desired to have empty slots. However, setting a high threshold means the preference lists will be larger, which could lead to an increase in the algorithm runtime. On the other hand, more placement options will be available for the tasks to be scheduled, reducing their wait time. In practice, the algorithm runtime may not necessarily increase. With more placement options available, the tasks may be scheduled sooner, thus leading to less tasks being scheduled per round, resulting in a decrease in the algorithm runtime per scheduling round. 

\begin{figure}
 \centering
        \includegraphics[width=0.4\textwidth]{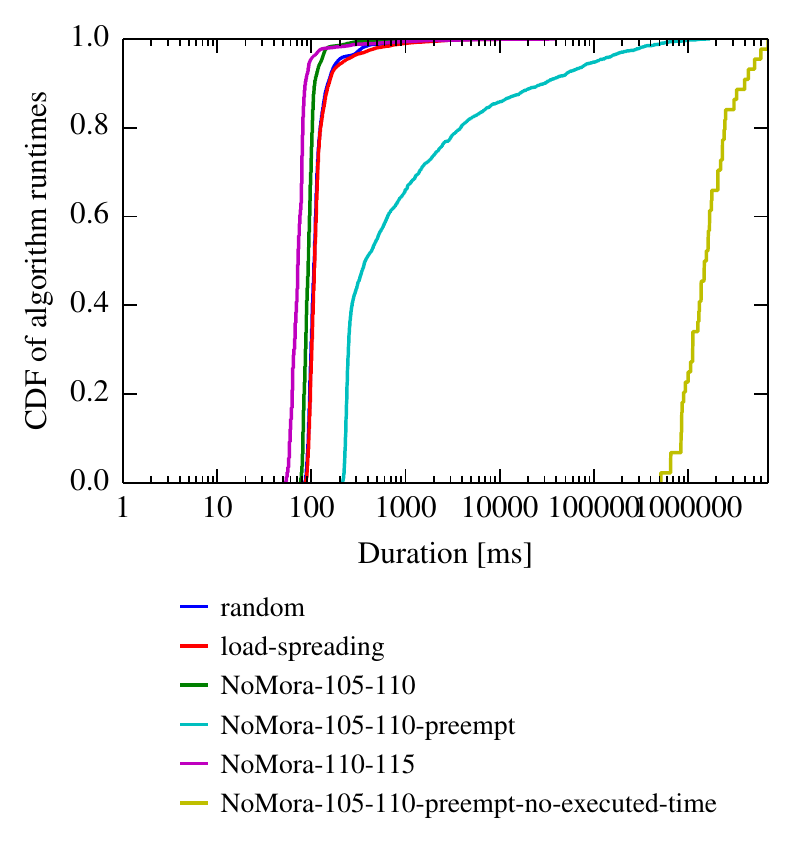}
        \caption{Algorithm runtime for different policies on the Google workload.}
       \label{fig:google-algorithm-runtime}
\end{figure}

Figure~\ref{fig:google-algorithm-runtime} presents results for the algorithm runtime for the load-spreading policy, random policy and for the NoMora policy with and without preemption on the Google workload. The two parameters of the cost model are set as in the previous experiment. 
The random and load-spreading policies have a similar algorithm runtime, with a median runtime of 108\millis-109\millis. However, the two policies differ at the tail: for the random policy the 99\thss percentile is 661\millis and a maximum of $18.89$s, while for load-spreading policy the 99\thss percentile is 974\millis and a maximum of $25.88$s. 
For NoMora with parameters $p_{m}=105$ and $p_{r}=110$, the median algorithm runtime is 93\millis (99\thss percentile is 248\millis, and maximum is $6.13$s),
an improvement of $1.61\times$ for the median runtime, and $2.66\times$ and $3.92\times$ at the 99\thss percentile, compared to the baselines.
For NoMora with $p_{m}=110$ and $p_{r}=115$, median runtime is 72\millis (99\thss percentile is 486\millis, and maximum is $39.55$s). 
On the other hand, the maximum value for the algorithm runtime is considerable larger in the case of NoMora policies.

\begin{figure}
 \centering
        \includegraphics[width=0.4\textwidth]{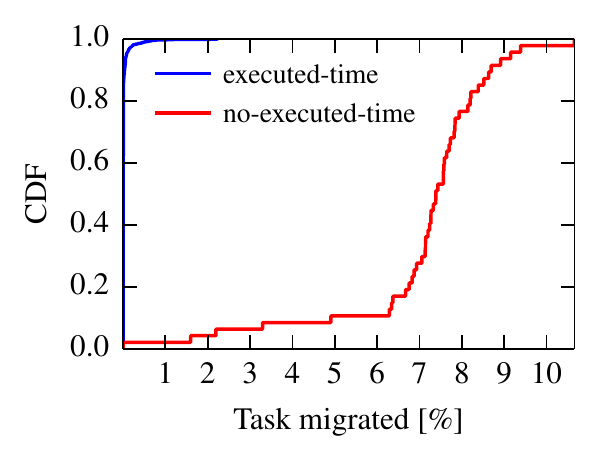}
        \caption{Percentage of migrated tasks for NoMora policy with preemption (parameters 105 and 110) on the Google workload.}
       \label{fig:google-percentage-migrated-tasks}
\end{figure}

NoMora with parameters $p_{m}=105$ and $p_{r}=110$ with preemption enabled 
takes a considerable longer amount of time, because of the higher number of arcs in the flow network compared to the case when preemption is not enabled (the arc preferences of the tasks that are running are not removed, unlike when preemption is not enabled), and the updates made to the flow graph (adding or changing running arcs to resources), further resulting in a larger number of tasks considered per scheduling round. This also translates into a larger task placement latency (\cref{sec:task-placement-latency}).
As can be seen from Figure~\ref{fig:google-percentage-migrated-tasks}, the percentage of migrated tasks in the first case (NoMora with preemption enabled and already executed time for a task considered in the arc computation) is on average $0.022$\% per scheduling round, with a 99\thss percentile of $0.5$\%. If $\beta_{i, j}=0$ (already executed time is not considered in the arc computation), a considerable number of task migrations take place: an average of $7.1$\% per scheduling round, with a 99\thss percentile of $10.07$\%. This happens because the time a task has already run is ignored in the arc cost computation, meaning that the cost is based solely on the expected application performance under the given network conditions.
The median algorithm runtime time is 373\millis, the 99\thss percentile is 511s and the maximum is $1719$s, which is $3.45\times$ larger than the baseline for the median runtime, and $773\times$ larger than the baseline at the 99\thss percentile. In the second case ($\beta_{i, j}=0$), the median running time is $1532$s, the 99\thss percentile $6610$s, and the maximum is $7118$s. This significant algorithm runtime means that preemption should be used with care. For example, only certain applications that explicitly demand to be migrated should be migrated, or migration can be triggered only if the application performance drops below a certain threshold. 

\subsection{Task placement latency}
\label{sec:task-placement-latency}

\begin{figure}
 \centering
        \includegraphics[width=0.4\textwidth]{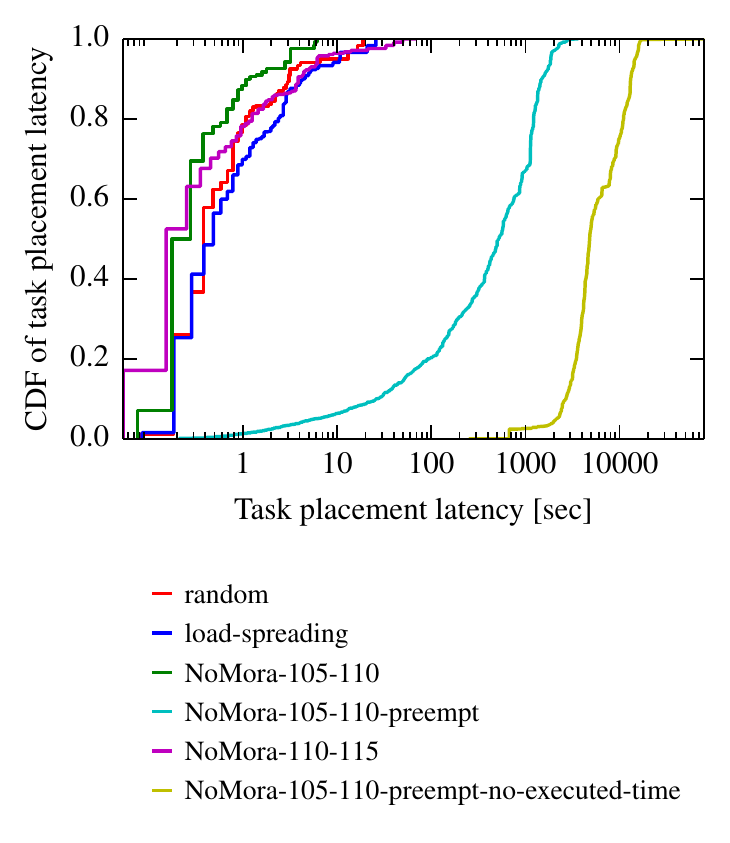}
        \caption{Task placement latency for different policies on the Google workload.}
       \label{fig:google-task-placement-delay}
\end{figure}

Figure~\ref{fig:google-task-placement-delay} presents the task placement latency, which is at the median 436\millis, 90\thss percentile is 312\millis and 99\thss percentile is 1.9s for the random policy; median is 498\millis, 90\thss percentile is 4.3s and 99\thss percentile is 25s for the load-spreading policy, median is 278\millis, 90\thss percentile is 1.23s and 99\thss percentile is 5.8s for the NoMora policy with parameters $p_{m}=105$ and $p_{r}=110$; median is 185\millis, 90\thss percentile is 1s and 99\thss percentile is 5.6s for the NoMora policy with parameters $p_{m}=110$ and $p_{r}=115$; 
median is 519s, 90\thss percentile is 1484s and 99\thss percentile is 2458s for the NoMora policy with preemption enabled and parameters $p_{m}=105$ and $p_{r}=110$; and
median is 4812s, 90\thss percentile is 13077s and 99\thss percentile is 16251s for the NoMora policy with preemption enabled, $\beta_{i, j}=0$ and parameters $p_{m}=105$ and $p_{r}=110$.

The NoMora policy with parameters $p_{m}=105$ and $p_{r}=110$ improves the median task placement latency by $1.56\times$ compared to the random policy and by $1.79\times$ compared to the load-spreading policy.
The NoMora policy with parameters $p_{m}=110$ and $p_{r}=115$ improves the median task placement latency by $2.35\times$ compared to the random policy and by $2.69\times$ compared to the load-spreading policy.

\begin{figure}
 \centering
        \includegraphics[width=0.4\textwidth]{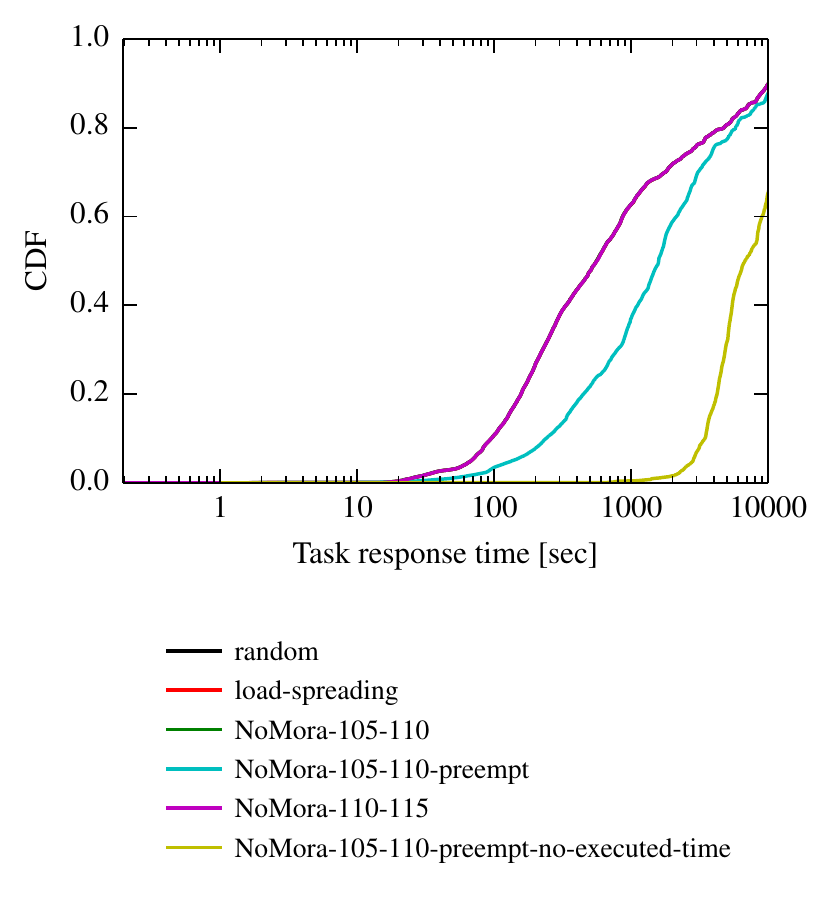}
        \caption{Task response time for different policies on the Google workload.}
       \label{fig:google-task-response-time}
\end{figure}

In Figure~\ref{fig:google-task-response-time}, 
it can be seen that the NoMora policy with preemption degrades the task response time, because of longer task placement latencies (Figure~\ref{fig:google-task-placement-delay}). The CDF is truncated to 10,000s, because the trace includes long-running jobs that span the whole trace. 

\section{Limitations}
\label{sec:limitations}


\textbf{Root task placement} Gang scheduling means all the application's tasks are placed simultaneously. Certain applications (\eg, Spark, Tensorflow, STRADS), whose computation is done by all tasks together would benefit from having all their tasks placed simultaneously, reducing the task wait time and overall job runtime. 
In the current NoMora policy, the root task is placed before the other tasks in any available slot in the cluster, and the optimal placement of the other tasks is computed based on the placement of the root task. This can negatively impact the overall job performance. With gang scheduling, the cost between the root task and the other tasks is not known in the beginning, making it difficult to decide where to place all the tasks from the start. 

\textbf{Changes in network traffic} 
NoMora accounts for the latency changes introduced by the traffic from tasks already placed in previous rounds. It does not account for the changes produced by tasks placed in the same round, \eg, the network latency measurements are not updated during the scheduling algorithm runtime. If a task's placement proves to be suboptimal due to the traffic generated by the tasks placed in the same scheduling round, the task can be migrated to a better placement. 


\textbf{Network bandwidth demands} NoMora can be extended to avoid bandwidth oversubcription at the end-host, or bandwidth allocation can be done by a different system (\cref{sec:bk-cluster-scheduler}). 

\section{Related work}
\label{sec:bk-cluster-scheduler}

In general, incorporating network demands within the cluster scheduler has been treated as a separate problem from cluster schedulers that take into account only the host resources required by a job. 

\textbf{Network bandwidth guarantees} Network throughput variability in cloud providers was an important issue, with bandwidth varying by a factor of five in some cases~\cite{ballani2011}, leading to uncertain  application performance and, consequently, tenant cost. 
Nowadays, cloud providers' commercial offerings list the expected network bandwidth for each type of VM. These changes fit with the observation that network bandwidth guarantees have improved in recent years~\cite{persico-ec2,persico-azure}. 

This improvement in network bandwidth guaranatees for tenants is the result of substantial research in the area.
Allocating network bandwidth between endpoints was first described in the context of Virtual Private Networks (VPN)~\cite{hose}. In the cloud computing model, the VPN customers can be assimilated with the tenants from the cloud, and a VPN endpoint's equivalent is a VM.
The customer-pipe model is the allocation of bandwidth on paths between source-destination pairs of endpoints of the VPN. In this model, a full mesh between customers is required to satisfy the SLAs. 
In the hose model, an endpoint is connected with a set of endpoints, but the bandwidth allocation is not specified between pairs. Instead, the aggregate bandwidth required for the outgoing traffic to the other endpoints and the aggregate bandwidth required for the incoming traffic from the other endpoints in the hose is specified.
These two models served as basis for bandwidth allocation in the cloud, with the hose model being frequently used~\cite{secondnet, gatekeeper, ballani2011, eyeq}. SecondNet~\cite{secondnet} introduces an abstraction called virtual data centre (VDC) and multiple types of services. 
Gatekeeper~\cite{gatekeeper} and EyeQ~\cite{eyeq} use the hose model~\cite{hose} and set minimum bandwidth guarantees for sending and receving traffic for a VM, which can be increased up to a maximum rate if unused capacity is available. 
ElasticSwitch~\cite{elastic} provides minimum bandwidth guarantees by dividing the hose model guarantees into VM-to-VM guarantees, and taking the minimum between the guarantees of the two VMs. 
Oktopus~\cite{ballani2011} uses the hose model (virtual cluster, where all the VMs are connected through a single switch) and virtual oversubscribed cluster model (groups of virtual clusters connected through a switch with an oversubscription factor). 
Proteus~\cite{proteus} authors analyse the traffic patterns of several MapReduce jobs, finding that the network demands of the applications change over time. They propose a time-varying network bandwidth allocation scheme, which is a variant of the hose model. 
CloudMirror~\cite{cloudmirror} derives a network abstraction model, tenant application graph, based on the application's communication pattern, which have multiple tiers or components, and bandwidth within the component is allocated using the hose model. 
Profiling applications to determine their network throughput can help allocate bandwidth in a more efficient manner~\cite{proteus, cicada}. This is similar to modeling the relationship between application performance and network latency based on experimental data, as we have done in Section~\ref{sec:impact-model}. 
Choreo~\cite{choreo} measures the network throughput between VM pairs through packet trains, estimates the cross traffic, and locates bottleneck links. Based on the network measurements and application profiles (number of bytes sent), it makes VM placement decisions to minimise application completion time.
The Choreo system is the closest system to the NoMora cluster scheduling architecture, where we use network latency measurements between pairs of hosts and application performance predictions dependent upon current network latency to decide where to place tenants' VMs. 
VM placement when providing network bandwidth guarantees usually starts by looking at subtrees in the topology to place the VMs, and goes upward in the tree to find a suitable allocation~\cite{ballani2011, proteus, ballani2013, cloudmirror}. 
CloudMirror~\cite{cloudmirror} additionally incorporates an anti-colocation constraint in the VM placement algorithm to ensure availability. SecondNet~\cite{secondnet} builds a bipartite graph whose nodes are the VMs on the left side and the physical machines on the right side, and then finds a matching based on the weights of the edges of the graph using min-cost max-flow. The weights are assigned based on the available bandwidth of the corresponding server. 
This approach is similar to Quincy~\cite{quincy} and Firmament~\cite{firmament}, which also model the scheduling problem as a min-cost max-flow problem. Firmament's network-aware policy avoids bandwidth oversubscription at the end-host by incorporating applications network bandwidth demands into the flow graph. \cite{infocomvm} proposes an algorithm for traffic-aware VM placement that takes into account the traffic rates between VMs, and studies how different traffic patterns and \dc network architectures impact the algorithm's outcome. 

\textbf{Tail network latency guarantees} 
Silo~\cite{silo} controls tenant's bandwidth to bound network queueing delay through packet pacing at the end-host. It then places VMs using a first-fit algorithm, while trying to place a tenant's VMs to minimise the amount of network traffic that the core links have to cary. QJump~\cite{qjump} computes rate limits 
for classes of applications, ranging from latency-sensitive applications for which it offeres strict latency guarantees to throughput-intensive ones for which latency can be variable. 
SNC-Meister~\cite{sncmeister} provides latency guarantees for lower percentiles, \eg, $99.9$\thss percentile, admitting more tenants in a \dc as a result.
It analyses tenant traces and computes their tail latencies, and then performs VM admission control.  
 

\section{Conclusion}

We introduced  latency-driven, application performance-aware cluster scheduling, and NoMora, a cluster scheduling framework that implements this type of policy. It exploits functions that predict application performance based upon network latency
and dynamic network latency measurements between hosts to place tasks in a \dc, providing them with improved application performance. The overall application performance improvement given by NoMora depends on the workload, network topology and on the network conditions in the \dc.
Using the Google cluster workload augmented with cloud latency measurements from~\cite{diana-tma} and with performance prediction functions, we showed that the NoMora policy improves the overall average application performance by up to $13.4$\% and by up to 42\% if preemption is enabled, and improves the task placement latency by a factor of $1.79\times$ and the median algorithm runtime by $1.16\times$ compared to the baselines. This demonstrates that application performance can be improved by exploiting the relationship between network latency and application performance, and the current network conditions in a \dc, while preserving the demands of low-latency cluster scheduling. 

In future work, we will evaluate our cluster scheduling architecture on different cluster workloads and on an experimental testbed.


\section{Acknowledgements}
We would like to thank Ionel Gog and Malte Schwarzkopf for helping with Firmament. We would like to thank Noa Zilberman, who developed the latency appliance used to conduct the measurements described in Section~\ref{sec:injecting}, and previously presented in~\cite{pam2017, diana-report}. The latency appliance was first published in~\cite{pam2017}, and was open sourced in March 2017 and can be found at~\cite{latency-gadget}.

\bibliographystyle{abbrv}
\bibliography{reference}

\begin{thebibliography}{10}

\bibitem{tensorflow}
M.~Abadi, P.~Barham, J.~Chen, Z.~Chen, A.~Davis, J.~Dean, M.~Devin,
  S.~Ghemawat, G.~Irving, M.~Isard, M.~Kudlur, J.~Levenberg, R.~Monga,
  S.~Moore, D.~G. Murray, B.~Steiner, P.~Tucker, V.~Vasudevan, P.~Warden,
  M.~Wicke, Y.~Yu, and X.~Zheng.
\newblock Tensorflow: A system for large-scale machine learning.
\newblock In {\em Proceedings of the 12th USENIX Conference on Operating
  Systems Design and Implementation}, OSDI'16, pages 265--283, Berkeley, CA,
  USA, 2016. USENIX Association.

\bibitem{netnorad}
A.~Adams, P.~Lapukhov, and J.~H. Zeng.
\newblock {NetNORAD: Troubleshooting networks via end-to-end probing}.
\newblock
  \url{https://code.facebook.com/posts/1534350660228025/netnorad-troubleshooting-networks-via-end-to-end-probing/},
  2016.
\newblock [Online; accessed December 2018].

\bibitem{vmm}
S.~Akoush, R.~Sohan, A.~Rice, A.~W. Moore, and A.~Hopper.
\newblock Predicting the performance of virtual machine migration.
\newblock In {\em 2010 IEEE International Symposium on Modeling, Analysis and
  Simulation of Computer and Telecommunication Systems}, pages 37--46, Aug
  2010.

\bibitem{fattree}
M.~Al-Fares, A.~Loukissas, and A.~Vahdat.
\newblock A scalable, commodity data center network architecture.
\newblock In {\em Proceedings of the ACM SIGCOMM 2008 Conference on Data
  Communication}, SIGCOMM '08, pages 63--74, New York, NY, USA, 2008. ACM.

\bibitem{Hedera}
M.~Al-Fares, S.~Radhakrishnan, B.~Raghavan, N.~Huang, and A.~Vahdat.
\newblock Hedera: Dynamic flow scheduling for data center networks.
\newblock In {\em Proceedings of the 7th USENIX Conference on Networked Systems
  Design and Implementation}, NSDI'10, pages 19--19, Berkeley, CA, USA, 2010.
  USENIX Association.

\bibitem{pfabric}
M.~Alizadeh, S.~Yang, M.~Sharif, S.~Katti, N.~McKeown, B.~Prabhakar, and
  S.~Shenker.
\newblock pfabric: Minimal near-optimal datacenter transport.
\newblock In {\em Proceedings of the ACM SIGCOMM 2013 Conference on SIGCOMM},
  SIGCOMM '13, pages 435--446, New York, NY, USA, 2013. ACM.

\bibitem{facebook-dc}
A.~Andreyev.
\newblock Introducing data center fabric, the next-generation facebook data
  center network.
\newblock
  \url{https://code.fb.com/production-engineering/introducing-data-center-fabric-the-next-generation-facebook-data-center-network/},
  2014.
\newblock [Online; accessed December 2018].

\bibitem{Atikoglu:2012}
B.~Atikoglu, Y.~Xu, E.~Frachtenberg, S.~Jiang, and M.~Paleczny.
\newblock Workload analysis of a large-scale key-value store.
\newblock In {\em Proceedings of the 12th ACM SIGMETRICS/PERFORMANCE Joint
  International Conference on Measurement and Modeling of Computer Systems},
  SIGMETRICS '12, pages 53--64, New York, NY, USA, 2012. ACM.

\bibitem{ballani2011}
H.~Ballani, P.~Costa, T.~Karagiannis, and A.~Rowstron.
\newblock Towards predictable datacenter networks.
\newblock In {\em Proceedings of the ACM SIGCOMM 2011 Conference}, SIGCOMM '11,
  pages 242--253, New York, NY, USA, 2011. ACM.

\bibitem{ballani2013}
H.~Ballani, K.~Jang, T.~Karagiannis, C.~Kim, D.~Gunawardena, and G.~O'Shea.
\newblock Chatty tenants and the cloud network sharing problem.
\newblock In {\em Proceedings of the 10th USENIX Conference on Networked
  Systems Design and Implementation}, nsdi'13, pages 171--184, Berkeley, CA,
  USA, 2013. USENIX Association.

\bibitem{barker:2010}
S.~K. Barker and P.~Shenoy.
\newblock Empirical evaluation of latency-sensitive application performance in
  the cloud.
\newblock In {\em Proceedings of the First Annual ACM SIGMM Conference on
  Multimedia Systems}, MMSys '10, pages 35--46, New York, NY, USA, 2010. ACM.

\bibitem{MicroTE}
T.~Benson, A.~Anand, A.~Akella, and M.~Zhang.
\newblock {MicroTE: Fine Grained Traffic Engineering for Data Centers}.
\newblock In {\em Proceedings of the Seventh COnference on Emerging Networking
  EXperiments and Technologies}, CoNEXT '11, pages 8:1--8:12, New York, NY,
  USA, 2011. ACM.

\bibitem{varys}
M.~Chowdhury, Y.~Zhong, and I.~Stoica.
\newblock Efficient coflow scheduling with varys.
\newblock In {\em Proceedings of the 2014 ACM Conference on SIGCOMM}, SIGCOMM
  '14, pages 443--454, New York, NY, USA, 2014. ACM.

\bibitem{spark-perf}
Databricks.
\newblock {Spark-perf benchmark}.
\newblock \url{https://github.com/databricks/spark-perf}, 2018.
\newblock [Online; accessed December 2018].

\bibitem{task}
F.~R. Dogar, T.~Karagiannis, H.~Ballani, and A.~Rowstron.
\newblock Decentralized task-aware scheduling for data center networks.
\newblock In {\em Proceedings of the 2014 ACM Conference on SIGCOMM}, SIGCOMM
  '14, pages 431--442, New York, NY, USA, 2014. ACM.

\bibitem{hose}
N.~G. Duffield, P.~Goyal, A.~Greenberg, P.~Mishra, K.~K. Ramakrishnan, and
  J.~E. van~der Merive.
\newblock A flexible model for resource management in virtual private networks.
\newblock In {\em Proceedings of the Conference on Applications, Technologies,
  Architectures, and Protocols for Computer Communication}, SIGCOMM '99, pages
  95--108, New York, NY, USA, 1999. ACM.

\bibitem{firmament}
I.~Gog, M.~Schwarzkopf, A.~Gleave, R.~N.~M. Watson, and S.~Hand.
\newblock Firmament: Fast, centralized cluster scheduling at scale.
\newblock In {\em Proceedings of the 12th USENIX Conference on Operating
  Systems Design and Implementation}, OSDI'16, pages 99--115, Berkeley, CA,
  USA, 2016. USENIX Association.

\bibitem{qjump}
M.~P. Grosvenor, M.~Schwarzkopf, I.~Gog, R.~N.~M. Watson, A.~W. Moore, S.~Hand,
  and J.~Crowcroft.
\newblock Queues don't matter when you can jump them!
\newblock In {\em Proceedings of the 12th USENIX Conference on Networked
  Systems Design and Implementation}, NSDI'15, pages 1--14, Berkeley, CA, USA,
  2015. USENIX Association.

\bibitem{secondnet}
C.~Guo, G.~Lu, H.~J. Wang, S.~Yang, C.~Kong, P.~Sun, W.~Wu, and Y.~Zhang.
\newblock Secondnet: A data center network virtualization architecture with
  bandwidth guarantees.
\newblock In {\em Proceedings of the 6th International COnference}, Co-NEXT
  '10, pages 15:1--15:12, New York, NY, USA, 2010. ACM.

\bibitem{pingmesh}
C.~Guo, L.~Yuan, D.~Xiang, Y.~Dang, R.~Huang, D.~Maltz, Z.~Liu, V.~Wang,
  B.~Pang, H.~Chen, Z.-W. Lin, and V.~Kurien.
\newblock Pingmesh: A large-scale system for data center network latency
  measurement and analysis.
\newblock In {\em Proceedings of the 2015 ACM Conference on Special Interest
  Group on Data Communication}, SIGCOMM '15, pages 139--152, New York, NY, USA,
  2015. ACM.

\bibitem{netem}
S.~Hemminger.
\newblock {NetEm - Network Emulator}.
\newblock \url{http://man7.org/linux/man-pages/man8/tc-netem.8.html}.
\newblock [Online; accessed December 2018].

\bibitem{quincy}
M.~Isard, V.~Prabhakaran, J.~Currey, U.~Wieder, K.~Talwar, and A.~Goldberg.
\newblock Quincy: Fair scheduling for distributed computing clusters.
\newblock In {\em Proceedings of the ACM SIGOPS 22Nd Symposium on Operating
  Systems Principles}, SOSP '09, pages 261--276, New York, NY, USA, 2009. ACM.

\bibitem{silo}
K.~Jang, J.~Sherry, H.~Ballani, and T.~Moncaster.
\newblock Silo: Predictable message latency in the cloud.
\newblock In {\em Proceedings of the 2015 ACM Conference on Special Interest
  Group on Data Communication}, SIGCOMM '15, pages 435--448, New York, NY, USA,
  2015. ACM.

\bibitem{eyeq}
V.~Jeyakumar, M.~Alizadeh, D.~Mazi\`{e}res, B.~Prabhakar, C.~Kim, and
  A.~Greenberg.
\newblock {EyeQ: Practical Network Performance Isolation at the Edge}.
\newblock In {\em Proceedings of the 10th USENIX Conference on Networked
  Systems Design and Implementation}, nsdi'13, pages 297--312, Berkeley, CA,
  USA, 2013. USENIX Association.

\bibitem{Kim:2016}
J.~K. Kim, Q.~Ho, S.~Lee, X.~Zheng, W.~Dai, G.~A. Gibson, and E.~P. Xing.
\newblock Strads: A distributed framework for scheduled model parallel machine
  learning.
\newblock In {\em Proceedings of the Eleventh European Conference on Computer
  Systems}, EuroSys '16, pages 5:1--5:16, New York, NY, USA, 2016. ACM.

\bibitem{strads}
J.~K. Kim, Q.~Ho, S.~Lee, X.~Zheng, W.~Dai, G.~A. Gibson, and E.~P. Xing.
\newblock {STRADS}.
\newblock \url{https://github.com/petuum/strads.git}, 2018.
\newblock [Online; accessed December 2018].

\bibitem{bwe}
A.~Kumar, S.~Jain, U.~Naik, A.~Raghuraman, N.~Kasinadhuni, E.~C. Zermeno, C.~S.
  Gunn, J.~Ai, B.~Carlin, M.~Amarandei-Stavila, M.~Robin, A.~Siganporia,
  S.~Stuart, and A.~Vahdat.
\newblock Bwe: Flexible, hierarchical bandwidth allocation for wan distributed
  computing.
\newblock In {\em Proceedings of the 2015 ACM Conference on Special Interest
  Group on Data Communication}, SIGCOMM '15, pages 1--14, New York, NY, USA,
  2015. ACM.

\bibitem{choreo}
K.~LaCurts, S.~Deng, A.~Goyal, and H.~Balakrishnan.
\newblock Choreo: Network-aware task placement for cloud applications.
\newblock In {\em Proceedings of the 2013 Conference on Internet Measurement
  Conference}, IMC '13, pages 191--204, New York, NY, USA, 2013. ACM.

\bibitem{cicada}
K.~LaCurts, J.~C. Mogul, H.~Balakrishnan, and Y.~Turner.
\newblock Cicada: Introducing predictive guarantees for cloud networks.
\newblock In {\em Proceedings of the 6th USENIX Conference on Hot Topics in
  Cloud Computing}, HotCloud'14, pages 14--14, Berkeley, CA, USA, 2014. USENIX
  Association.

\bibitem{mnist}
Y.~LeCun and C.~Cortes.
\newblock {MNIST} handwritten digit database.
\newblock \url{http://yann.lecun.com/exdb/mnist/}, 2010.
\newblock [Online; accessed December 2018].

\bibitem{cloudmirror}
J.~Lee, Y.~Turner, M.~Lee, L.~Popa, S.~Banerjee, J.-M. Kang, and P.~Sharma.
\newblock Application-driven bandwidth guarantees in datacenters.
\newblock In {\em Proceedings of the 2014 ACM Conference on SIGCOMM}, SIGCOMM
  '14, pages 467--478, New York, NY, USA, 2014. ACM.

\bibitem{mutilate}
J.~Leverich.
\newblock {Mutilate: high-performance memcached load generator}.
\newblock \url{https://github.com/leverich/mutilate}, 2014.
\newblock [Online; accessed December 2018].

\bibitem{memcached}
Memcached.
\newblock {Memcached}.
\newblock \url{https://memcached.org/}, 2018.
\newblock [Online; accessed December 2018].

\bibitem{infocomvm}
X.~Meng, V.~Pappas, and L.~Zhang.
\newblock {Improving the Scalability of Data Center Networks with Traffic-aware
  Virtual Machine Placement}.
\newblock In {\em Proceedings of the 29th Conference on Information
  Communications}, INFOCOM'10, pages 1154--1162, Piscataway, NJ, USA, 2010.
  IEEE Press.

\bibitem{mogul:2015}
J.~C. Mogul and R.~R. Kompella.
\newblock Inferring the network latency requirements of cloud tenants.
\newblock In {\em Proceedings of the 15th USENIX Conference on Hot Topics in
  Operating Systems}, HOTOS'15, pages 24--24, Berkeley, CA, USA, 2015. USENIX
  Association.

\bibitem{fastpass}
J.~Perry, A.~Ousterhout, H.~Balakrishnan, D.~Shah, and H.~Fugal.
\newblock Fastpass: A centralized "zero-queue" datacenter network.
\newblock In {\em Proceedings of the 2014 ACM Conference on SIGCOMM}, SIGCOMM
  '14, pages 307--318, New York, NY, USA, 2014. ACM.

\bibitem{persico-ec2}
V.~Persico, P.~Marchetta, A.~Botta, and A.~Pescap\`{e}.
\newblock Measuring network throughput in the cloud.
\newblock {\em Comput. Netw.}, 93(P3):408--422, Dec. 2015.

\bibitem{persico-azure}
V.~Persico, P.~Marchetta, A.~Botta, and A.~Pescape.
\newblock On network throughput variability in microsoft azure cloud.
\newblock In {\em 2015 IEEE Global Communications Conference (GLOBECOM)}, pages
  1--6, Dec 2015.

\bibitem{elastic}
L.~Popa, P.~Yalagandula, S.~Banerjee, J.~C. Mogul, Y.~Turner, and J.~R. Santos.
\newblock Elasticswitch: Practical work-conserving bandwidth guarantees for
  cloud computing.
\newblock In {\em Proceedings of the ACM SIGCOMM 2013 Conference on SIGCOMM},
  SIGCOMM '13, pages 351--362, New York, NY, USA, 2013. ACM.

\bibitem{diana-thesis}
D.~A. Popescu.
\newblock {Latency-driven performance in data centres}.
\newblock Technical Report UCAM-CL-TR-937, University of Cambridge, Computer
  Laboratory, June 2019.

\bibitem{diana-mascots}
D.~A. Popescu and A.~W. Moore.
\newblock {PTPmesh: Data Center Network Latency Measurements Using PTP}.
\newblock In {\em 2017 IEEE 25th International Symposium on Modeling, Analysis,
  and Simulation of Computer and Telecommunication Systems (MASCOTS)}, pages
  73--79, Sept 2017.

\bibitem{diana-tma}
D.~A. Popescu and A.~W. Moore.
\newblock {A First Look At Data Center Network Conditions Through The Eyes of
  PTPmesh}.
\newblock In {\em Proceedings of the 2018 IFIP/IEEE 2nd Network Traffic
  Measurement and Analysis Conference (TMA 2018)}, TMA '18. IFIP, 2018.

\bibitem{diana-report}
D.~A. Popescu, N.~Zilberman, and A.~W. Moore.
\newblock {Characterizing the impact of network latency on cloud-based
  applications' performance}.
\newblock Technical Report UCAM-CL-TR-914, University of Cambridge, Computer
  Laboratory, Nov. 2017.

\bibitem{google-workload}
C.~Reiss, A.~Tumanov, G.~R. Ganger, R.~H. Katz, and M.~A. Kozuch.
\newblock Heterogeneity and dynamicity of clouds at scale: Google trace
  analysis.
\newblock In {\em Proceedings of the Third ACM Symposium on Cloud Computing},
  SoCC '12, pages 7:1--7:13, New York, NY, USA, 2012. ACM.

\bibitem{gatekeeper}
H.~Rodrigues, J.~R. Santos, Y.~Turner, P.~Soares, and D.~Guedes.
\newblock Gatekeeper: Supporting bandwidth guarantees for multi-tenant
  datacenter networks.
\newblock In {\em Proceedings of the 3rd Conference on I/O Virtualization},
  WIOV'11, pages 6--6, Berkeley, CA, USA, 2011. USENIX Association.

\bibitem{vnet-pingmesh}
A.~Roy, D.~Bansal, D.~Brumley, H.~K. Chandrappa, P.~Sharma, R.~Tewari,
  B.~Arzani, and A.~C. Snoeren.
\newblock Cloud datacenter sdn monitoring: Experiences and challenges.
\newblock In {\em Proceedings of the Internet Measurement Conference 2018}, IMC
  '18, pages 464--470, New York, NY, USA, 2018. ACM.

\bibitem{scipy}
Scipy.
\newblock Scipy.
\newblock \url{https://www.scipy.org/}.
\newblock [Online; accessed December 2018].

\bibitem{spark}
A.~Spark.
\newblock {Apache Spark MLLib}.
\newblock \url{https://spark.apache.org/}.
\newblock [Online; accessed December 2018].

\bibitem{lasso}
R.~Tibshirani.
\newblock Regression shrinkage and selection via the lasso: a retrospective.
\newblock {\em Journal of the Royal Statistical Society: Series B (Statistical
  Methodology)}, 73(3):273--282, 2011.

\bibitem{wang:2010}
G.~Wang and T.~S.~E. Ng.
\newblock The impact of virtualization on network performance of amazon ec2
  data center.
\newblock In {\em Proceedings of the 29th Conference on Information
  Communications}, INFOCOM'10, pages 1163--1171, Piscataway, NJ, USA, 2010.
  IEEE Press.

\bibitem{proteus}
D.~Xie, N.~Ding, Y.~C. Hu, and R.~Kompella.
\newblock The only constant is change: Incorporating time-varying network
  reservations in data centers.
\newblock In {\em Proceedings of the ACM SIGCOMM 2012 Conference on
  Applications, Technologies, Architectures, and Protocols for Computer
  Communication}, SIGCOMM '12, pages 199--210, New York, NY, USA, 2012. ACM.

\bibitem{bobtail}
Y.~Xu, Z.~Musgrave, B.~Noble, and M.~Bailey.
\newblock Bobtail: Avoiding long tails in the cloud.
\newblock In {\em Proceedings of the 10th USENIX Conference on Networked
  Systems Design and Implementation}, nsdi'13, pages 329--342, Berkeley, CA,
  USA, 2013. USENIX Association.

\bibitem{sncmeister}
T.~Zhu, D.~S. Berger, and M.~Harchol-Balter.
\newblock Snc-meister: Admitting more tenants with tail latency slos.
\newblock In {\em Proceedings of the Seventh ACM Symposium on Cloud Computing},
  SoCC '16, pages 374--387, New York, NY, USA, 2016. ACM.

\bibitem{latency-gadget}
N.~Zilberman.
\newblock {Latency Gadget}.
\newblock
  \url{https://github.com/NetFPGA/NetFPGA-SUME-public/wiki/Contrib-Project:-Latency-Gadget:-delay\_mb},
  2017.

\bibitem{pam2017}
N.~Zilberman, M.~Grosvenor, D.~A. Popescu, N.~Manihatty-Bojan, G.~Antichi,
  M.~Wojcik, and A.~W. Moore.
\newblock Where has my time gone?
\newblock In {\em Proceedings of the 18th International Conference on Passive
  and Active Measurement}, PAM 2017, 2017.

\end{thebibliography}

\end{document}